\documentclass[pdflatex,sn-nature]{sn-jnl}

\usepackage{graphicx}%
\usepackage{multirow}%
\usepackage{amsmath,amssymb,amsfonts}%
\usepackage{amsthm}%
\usepackage{mathrsfs}%
\usepackage[title]{appendix}%
\usepackage{xcolor}%
\usepackage{textcomp}%
\usepackage{manyfoot}%
\usepackage{booktabs}%
\usepackage{algorithm}%
\usepackage{algorithmicx}%
\usepackage{algpseudocode}%
\usepackage{listings}%
\usepackage{gensymb}%
\usepackage{upgreek}%
\usepackage{subfig}
\usepackage{subcaption}
\usepackage{float}

\begin{document}

\title[Design and performance of the coded mask for the Lunar Electromagnetic Monitor in X-rays (LEM-X)]{Design and performance of the coded mask for the Lunar Electromagnetic Monitor in X-rays (LEM-X)}


\author*[1,2]{\fnm{Yuri} \sur{Evangelista}}\email{yuri.evangelista@inaf.it}

\author[1,2]{\fnm{Alessio} \sur{Nuti}}

\author[1,2]{\fnm{Francesco} \sur{Ceraudo}}
\author[1,3]{\fnm{Edoardo} \sur{Giancarli}}
\author[4,5]{\fnm{Giuseppe} \sur{Dilillo}}
\author[6,7]{\fnm{Riccardo} \sur{Campana}}
\author[1,2]{\fnm{Giovanni} \sur{Della Casa}}
\author[1,2]{\fnm{Ettore} \sur{Del Monte}}
\author[1,2]{\fnm{Marco} \sur{Feroci}}
\author[8]{\fnm{Mauro} \sur{Fiorini}}
\author[1,2]{\fnm{Giovanni} \sur{Lombardi}}
\author[1]{\fnm{Massimo} \sur{Rapisarda}}
\author[9]{\vspace{0.15cm} \linebreak \fnm{\\Francesca} \sur{Esposito}}
\author[5]{\fnm{Immacolata} \sur{Donnarumma}}
\author[5]{\fnm{Alessandro} \sur{Turchi}}
\author[10]{\fnm{Ugo} \sur{Cortesi}}
\author[11]{\fnm{Fabio} \sur{D'Amico}}
\author[10]{\fnm{Marco} \sur{Gai}}
\author[11]{\fnm{Andrea} \sur{Argan}}

\affil*[1]{\orgdiv{INAF}, \orgname{Istituto di Astrofisica e Planetologia Spaziali}, \orgaddress{\street{Via del Fosso del Cavaliere 100}, \city{Rome}, \postcode{00133}, \country{Italy}}}
\affil[2]{\orgdiv{INFN}, \orgname{Sez. Roma Tor Vergata}, \orgaddress{\street{Via della Ricerca Scientifica}, \city{Rome}, \postcode{00133}, \country{Italy}}}
\affil[3]{\orgdiv{University of Rome ``La Sapienza''}, \orgname{Dept. of Physics}, \orgaddress{\street{Piazzale Aldo Moro 2}, \city{Rome}, \postcode{00185}, \country{Italy}}}

\affil[4]{\orgdiv{INAF}, \orgname{Osservatorio Astronomico di Roma}, \orgaddress{\street{Via Frascati 33}, \city{Rome}, \postcode{00078}, \country{Italy}}}
\affil[5]{\orgdiv{Agenzia Spaziale Italiana}, \orgaddress{\street{Via del Politecnico}, \city{Roma}, \postcode{00133}, \country{Italy}}}
\affil[6]{\orgdiv{INAF}, \orgname{Osservatorio di Astrofisica e Scienza dello Spazio}, \orgaddress{\street{Via Piero Gobetti 101}, \city{Bologna}, \postcode{40129}, \country{Italy}}}

\affil[7]{\orgdiv{INFN}, \orgname{Sez. Bologna}, \orgaddress{\street{Viale Berti Pichat 6/2}, \city{Bologna}, \postcode{40127}, \country{Italy}}}
\affil[8]{\orgdiv{INAF}, \orgname{Istituto di Astrofisica Spaziale e Fisica cosmica}, \orgaddress{\street{Via Alfonso Corti 12}, \city{Milano}, \postcode{20133}, \country{Italy}}}
\affil[9]{\orgdiv{INAF}, \orgname{Osservatorio Astronomico di Capodimonte}, \orgaddress{\street{Salita Moiariello 16}, \city{Napoli}, \postcode{80131}, \country{Italy}}}
\affil[10]{\orgdiv{CNR}, \orgname{Istituto di Fisica
Applicata “Nello Carrara”}, \orgaddress{\street{Via Madonna del Piano 10}, \city{Sesto Fiorentino}, \postcode{50019}, \country{Italy}}}
\affil[11]{\orgdiv{INAF}, \orgaddress{\street{Via del Parco Mellini 84}, \city{Roma}, \postcode{00136}, \country{Italy}}}

\abstract{The Lunar Electromagnetic Monitor in X-rays (LEM-X) is a proposed wide-field X-ray observatory designed for deployment on the Moon’s surface. Its primary scientific goal is to enhance multi-messenger astrophysics by detecting, localizing, and monitoring high-energy transient phenomena and variable X-ray sources across the sky. Building on the heritage of the eXTP and LOFT mission proposals, LEM-X employs pairs of coded-aperture cameras equipped with large-area linear Silicon Drift Detectors (SDDs), offering excellent spectral resolution ($\leq$350 eV at 6 keV) over the 2--50~keV energy range. Each camera provides a field of view of $\sim$1 steradian at 25\% effective area and achieves a Point-Source Location Accuracy (PSLA) of $\sim$1 arcminute, with an on-axis sensitivity better than 5~mCrab in 50~ks and $\sim$700~mCrab in 1 s. In this paper we describe the experiment and focus on the detailed design and optimization of the LEM-X coded mask, analyzing its scientific performance, imaging capabilities, and thermo-mechanical properties. We describe the mask code generation, decoding algorithms, and the trade-offs involved in achieving the required angular resolution, sensitivity, and structural integrity. Imaging simulations and mechanical analyses confirm the effectiveness of the proposed design, demonstrating its suitability for high-precision, wide-field X-ray imaging devoted to multi-messenger astrophysics and transient events detection.
}




\maketitle

\section{Introduction}\label{intro}

In the last years astrophysics has entered a new era with the detection of gravitational waves (GWs) \cite{Abbott2017} and the observation of their electromagnetic counterparts \cite{Troja2017, Goldstein2017}, as well as the discovery of high-energy neutrinos of cosmic origin \cite{IceCube2013}. This marks a shift toward multi-messenger astronomy, a novel approach that integrates observations from a variety of cosmic messengers including photons, GWs, neutrinos, and cosmic rays.
Since each messenger is produced by a distinct physical process, they carry unique information about the mechanisms within its source. Many of the most significant sources in multi-messenger astronomy are transient or variable, such as supernovae and Gamma-Ray Bursts (GRBs).
In this framework, the \emph{Lunar Electromagnetic Monitor in X-ray} (LEM-X) experiment \cite{DelMonte2024, DelMonte2025} is uniquely positioned to complement GW and neutrino observatories by providing fast X-ray localization and long-term monitoring data for a wide range of cosmic events. It will contribute to a more comprehensive understanding of the universe by combining observations from multiple cosmic messengers.
Moreover, thanks to its sensitivity and large field of view, LEM-X will play an important role in multi-wavelength astronomy, detecting and providing photon-by-photon data of transient and variable sources typical of the energetic Universe.

LEM-X is a wide-field, coded-aperture imaging telescope that operates in the 2--50~keV energy range. The observatory will be installed on the Lunar surface to simultaneously observe half of the sky, taking advantage of the Moon's stable environment and rotation to continuously monitor a large portion of the sky.

The project is developed within the context of the Earth-Moon-Mars (EMM) program, a strategic initiative aimed at developing a versatile infrastructure to support future exploration missions to the Moon and Mars. Led by the Italian National Institute for Astrophysics (INAF), in collaboration with the Italian Space Agency (ASI) and the National Research Council (CNR), EMM is funded under Italy’s National Recovery and Resilience Plan (NRRP), reflecting a national commitment to advance space science and exploration capabilities. By establishing a permanent presence on the lunar surface, the project envisions leveraging Moon's  unique vantage point for advanced astronomical observations of both Earth and Universe. Simultaneously, the lunar infrastructure will serve as a critical testbed for technologies and operational frameworks essential for the eventual human exploration of Mars.

\section{LEM-X instrument design}\label{sec:instrument_design}
\begin{table}[h!]
    \caption{LEM-X scientific requirements}\label{tab:sci_reqs}%
    \begin{tabular*}{\textwidth}{@{\extracolsep\fill}llll@{}}
        \toprule
        Requirement & Value  & Note\\
        \midrule
Field of View                   &	$\geq$5 sr\footnotemark[1]	                       &		\\
Energy band                     &	2--50~keV	                                       &		\\
Energy resolution               &	$\leq$350~eV FWHM at 6 keV	                       &		\\
Time resolution                 &	10~$\upmu$s	                                       &		\\
\multirow{2}{*}{Sensitivity}	&	$\leq$ 1000 mCrab\footnotemark[2]                  &	5$\sigma$, 1~s exposure	\\
	                            &	$\leq$ 5 mCrab\footnotemark[2]                     &	5$\sigma$, 1~ks exposure	\\
Angular resolution	            &	$\leq$ 5$' \times$5$'$ FWHM  	                   &		\\
Point source location accuracy	&	$\leq$ 1$' \times$1$'$ at 90$\%$ confidence level  &		\\

        \botrule
    \end{tabular*}
\footnotetext[1]{Considering a maximum Field of View (FoV) obstruction of  10$^\circ$ above the horizon, accounting for both the lunar platform and local terrain.}   
\footnotetext[2]{1 mCrab in the 2--50 keV energy range corresponds to 4.5$\times10^{-11}$ erg cm$^{-2}$ s$^{-1}$.}
\end{table}

\begin{figure}[t]
\centering
\includegraphics[width=0.9\textwidth]{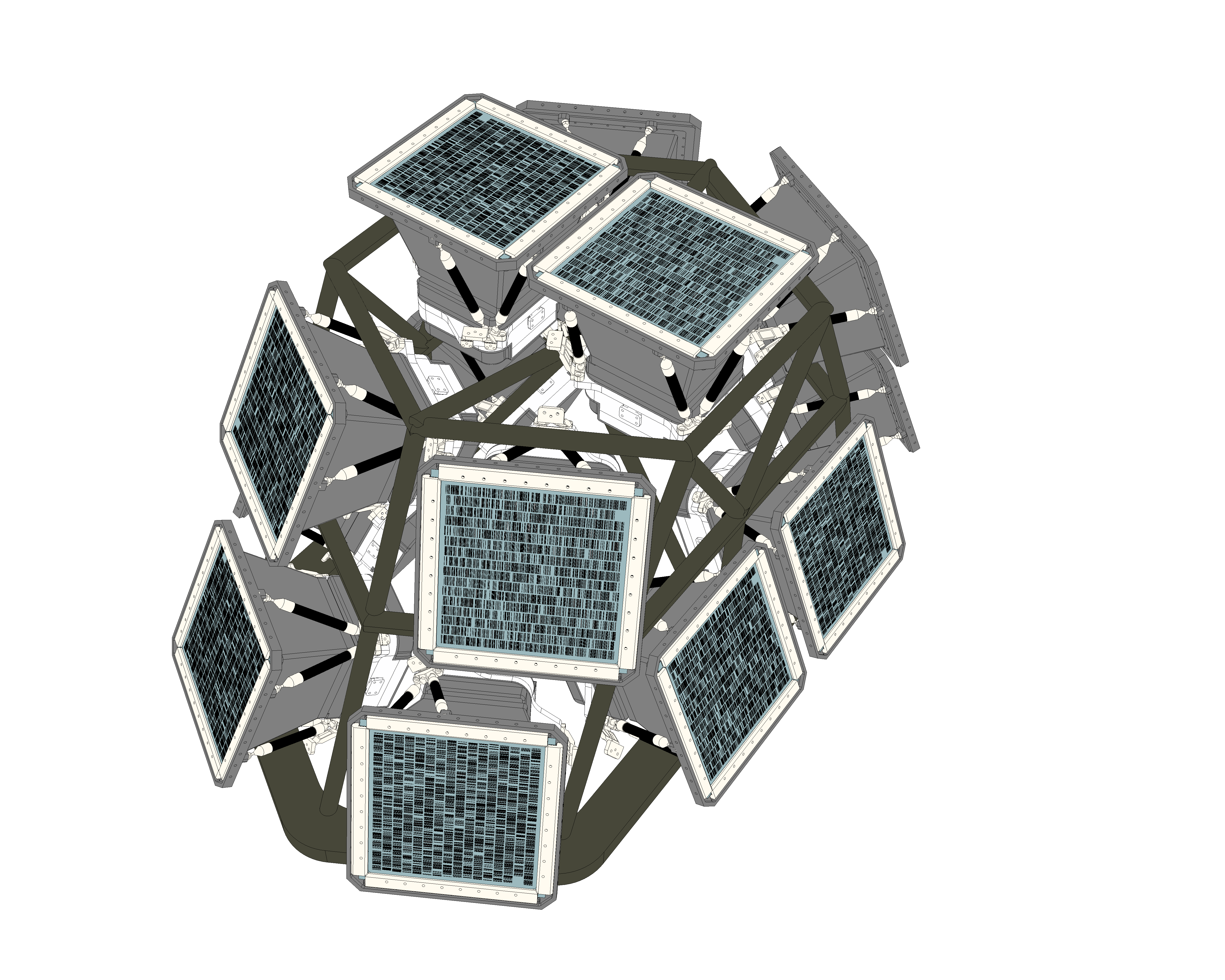} 
\caption{LEM-X baseline instrument design}\label{LEMX_fig}
\end{figure}

As reported in Table~\ref{tab:sci_reqs}, the instrument is designed to achieve an overall sensitivity better than 5 mCrab in 50 ks and 1 Crab (4.5$\times10^{-8}$ erg cm$^{-2}$ s$^{-1}$ in 2--50~keV) in 1 second in a 5~sr field of view (FoV), making it highly effective for detecting and characterizing transient events such as GRBs  and X-ray bursts. In addition, the large FoV will also allow for the long-term study of variable astrophysical sources.
The instrument is composed of several pairs of coded-aperture cameras. Each camera pair (hereafter a ``Unit'') uses orthogonally oriented detection planes and masks to achieve a Point-Source Location Accuracy of approximately 1 arcminute. This modular design provides an instantaneous field of view of about $90^\circ\times90^\circ$ ($\sim$2~steradians) per camera pair at zero response, with a single camera sensitivity better than 5 mCrab in 50 ks in a FoV $\geq 30^\circ \times 30^\circ$.
The baseline concept of the LEM-X observatory comprises 7 different camera pairs (Figure~\ref{LEMX_fig}), for a total of 14 cameras. One Unit is pointed to the zenith, while the others are equally spaced in azimuth with constant elevation angles. The homogeneity of the instrument effective area and sensitivity has been maximized considering $24^\circ$ of elevation for the 12 azimuthal cameras, as reported in \cite{Nuti2024} and \cite{DelMonte2024}.

\begin{figure}[h]
\centering
\includegraphics[width=2.25 in]{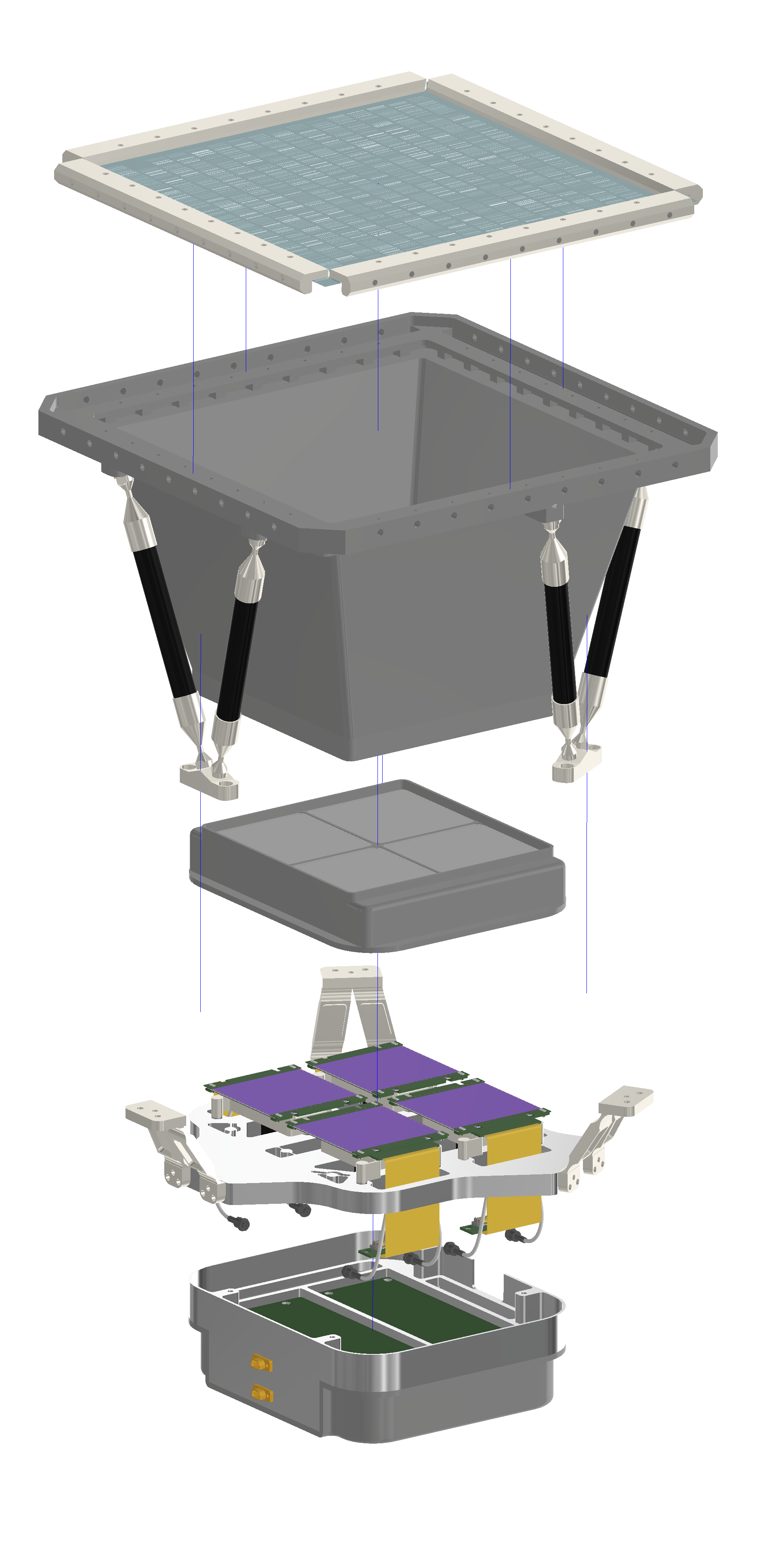}\\
\caption{Exploded view of the LEM-X camera. From top to bottom: coded mask, collimator and support structure, beryllium or polypropylene shield, detector plane, BEE box. Violet regions in the detector plane represent the active detector area.}
\end{figure}\label{fig:exploded_view}

The design of the LEM-X cameras is inherited from the \emph{enhanced X-ray Timing and Polarimetry mission} (eXTP)  WFM \cite{Hernanz2024}, and on the concept originally proposed for LOFT (\emph{Large Observatory For x-ray Timing}), which underwent a phase A study by ESA as M3 candidate \cite{Brandt2014}.

The cameras use the large area, linear Silicon Drift Detectors (SDDs) \cite{Rachevski2014J,Ceraudo2024SDD}. This technology, combined with low-noise read-out electronics \cite{Mele2025a}, provides excellent spectral-timing performance, with a Beginning-of-Life (BoL) energy resolution at room temperature better than 350~eV FWHM at 6~keV and a time resolution of 10~$\upmu$s.
Each camera includes one detector tray composed of four detector assemblies (DAs) \cite{Ceraudo2024SDD}, a beryllium or polypropylene layer acting as micrometeorite and orbital debris shield for the SDDs, a coded mask assembly with the mask covered by a thermal foil, a collimator, and a back-end electronics assembly (BEE). An exploded view of the LEM-X camera is shown in Figure~\ref{fig:exploded_view}.

Each DA features a 45.5~cm$^2$ sensitive area, 450~$\upmu$m thick SDD, the corresponding front-end readout electronics (FEE) and a mechanical support. The FEE includes 24 NOVA ASICs \cite{Mele2025a} (12 per SDD side). Two additional ASICs dedicated to the Analogue to Digital Conversion (ADC) of the electronic signals produced by each DA, are located in the BEE box, which also includes a power supply unit (PSU). 

\begin{figure}[t!]
\centering
\includegraphics[width=0.65\textwidth]{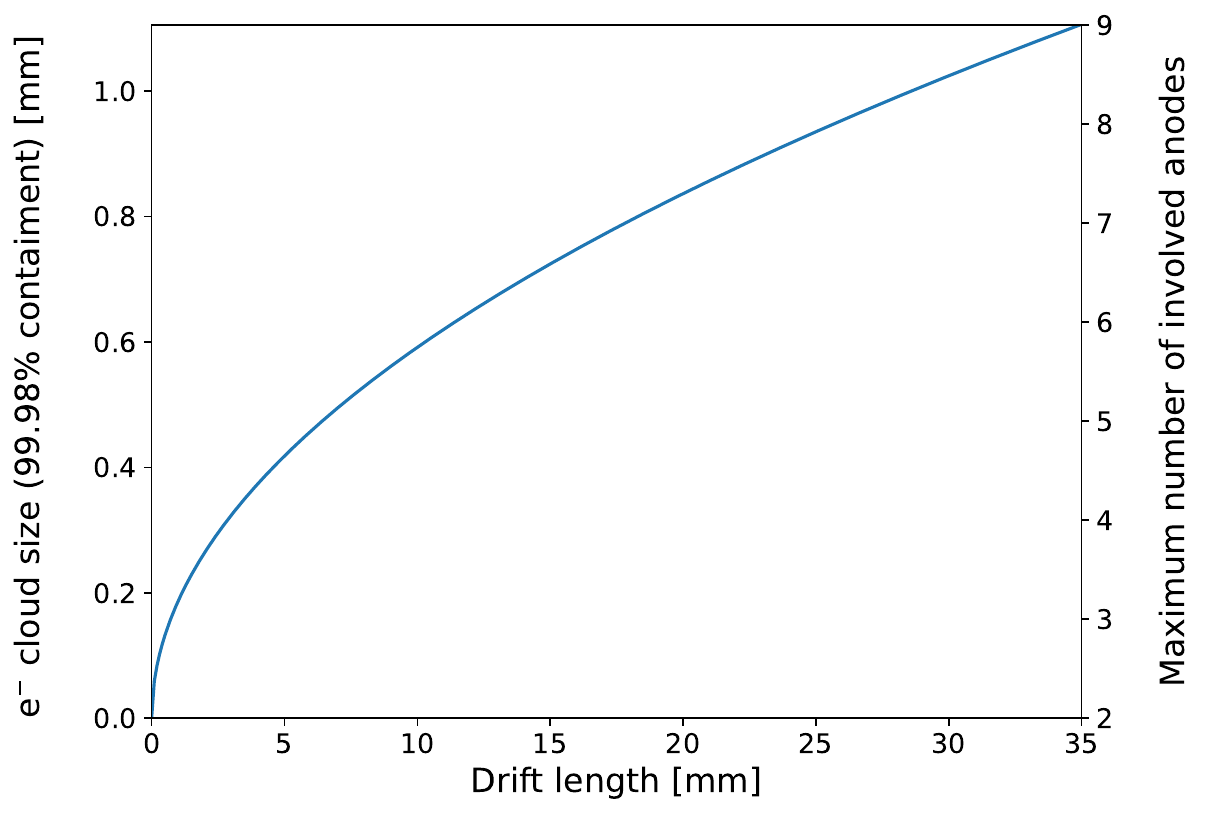}
\caption{Number of involved read-out anodes as a function of the charge cloud drift length, simulated at T$_{\mathrm{room}} = 20~^{\circ}\mathrm{C}$}\label{fig:involved_anodes}
\end{figure}

X-ray photons absorbed in the linear SDD create a cloud of electron-hole pairs. Under the action of a drift field ($\sim$360~V/cm) the electrons travel to the middle detector plane and then drift towards the closest collecting anodes. During the drift path, the electron cloud size increases as a result of the diffusion, so multiple anodes may read-out portions of the same charge cloud. The number of anodes needed to read-out the whole charge cloud for different drift lengths is shown in Figure~\ref{fig:involved_anodes}.

By simultaneously reading-out the charge collected by a large number ($\geq9$) of fine-pitched anodes (169~$\upmu$m pitch) it is possible to determine not only the energy deposited by the impinging photon, but also the bi-dimensional position of the photon interaction \cite{Campana2011,Ceraudo2024SDD}.
This technique provides a very good spatial resolution (better than 70~$\upmu$m FWHM) along the anode direction, while a rougher spatial accuracy is obtained along the drift direction ($<$8~mm at 95\% confidence level, 3~keV). For further details see \cite{Evangelista2014,Evangelista2012,Zwart2022, Ceraudo2024SDD}.

\section{Coded aperture imaging}\label{CAI}

\begin{figure}[t!]
\centering
\includegraphics[width=\textwidth]{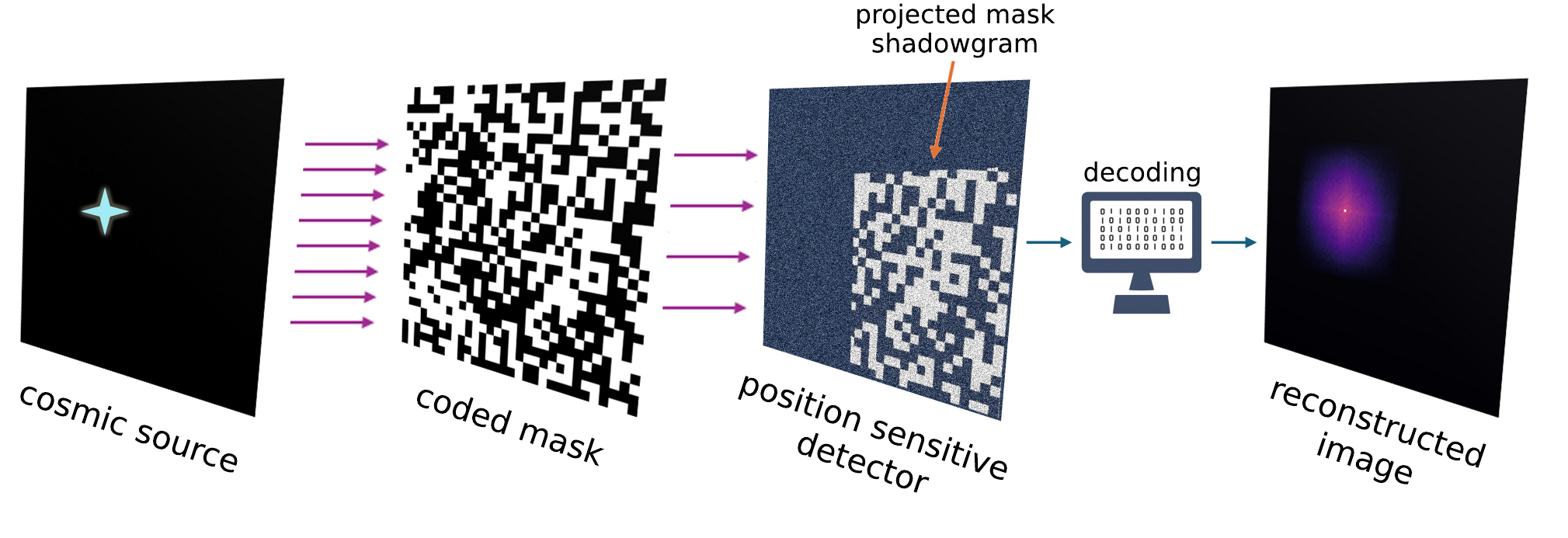}
\caption{Schematic representation of the coded aperture imaging (CAI) working principle. See Section~\ref{CAI} for more details.}\label{fig:cai_woriking_principle}
\end{figure}

As reported above, the design of the LEM-X cameras is based on the coded aperture imaging (CAI) concept. The CAI technique (Figure~\ref{fig:cai_woriking_principle}) exploits a layer of absorbing material with a known pattern of transparent and opaque areas (i.e., a coded mask) placed in front of a position sensitive detector. As radiation from a source passes through the mask open areas, it casts a unique shadow or ``shadowgram'' on the detector. The pattern and position of this shadow are a direct function of the source location and intensity.
Here below we will introduce the fundamental equations used in CAI. For comprehensive reviews on coded aperture techniques the reader is referred to \cite{Caroli1987,intZand1992,Accorsi2001,Braga2020,Goldwurm2022}.

\subsection{Decoding images from coded aperture instruments}\label{sec:decoding}

By using the balanced correlation method introduced by \cite{FenimoreCannon1978}, and taking into account the detector layout and any possible non-uniformity present in the detector pixel response (see, e.g., \cite{Goldwurm1995}), it is possible to deconvolve the recorded detector image to obtain the image of the observed sky (hereafter ``sky image'') in the whole field-of-view (FoV).
Let us consider the following 2-dimensional arrays:
\begin{itemize}
    \item  $D$: representing the detector image (i.e., the mask shadow cast by all the sources in the camera field-of-view);
    \item $R$: representing the decoding array, constructed from the mask pattern ($M$) using, for example, methods proposed by \cite{Caroli1987} and \cite{SkinnerPonman1994};
    \item $U$: representing the detector efficiency expressed as a number between 0 and 1 for each detector pixel, where $U_{ij}=0$ indicates a non-sensitive detector pixel (i.e., switched-off, not working or not physically present) and $U_{ij}=1$ indicates a detector pixel nominally working. Intermediate values (between 0 and 1) can be used, for example, for detector elements with a reduced collecting efficiency due to physical, geometric or electronic effects.
\end{itemize}

Then, the elements of the balanced reconstructed sky image ($S^\mathrm{bal}_{ij}$) can be calculated using:
\\
\begin{equation}\label{eq:deconvolution}
S^\mathrm{bal}_{ij} = (R \star D)_{ij} - (R \star U)_{ij} \cdot \frac{\sum D}{\sum U}
\end{equation}
where $\star$ represents the 2D aperiodic cross-correlation operator and $\sum$ the sum over all elements of the corresponding array. 

The second term in Eq. \ref{eq:deconvolution} corresponds to a flat-fielding operation, which corrects for non-uniformities in the detector response (represented by $U$) and normalizes the overall signal to account for variations in the total detected flux ($\Sigma D$) relative to the nominal total detector sensitivity ($\Sigma U$). 
The need for the \emph{flat-fielding} term can be easily understood considering, for example, the detector image produced by a flat background $B$
\begin{equation}
    D = B \odot U
\end{equation}
where $\odot$ represents the element-wise (Hadamard) product. 

Then, the resulting sky image is
\begin{equation}
   S^\mathbf{raw} = R \star D = R \star \left( B \odot U \right) = \bar{b} \cdot \left( R \star U \right)
\end{equation}
with $\bar{b}$ being the mean flat background counts collected by each detector pixel, or $\bar{b} = \Sigma D / \Sigma U$.

In presence of a flat background, we want our balanced sky image $S^\mathrm{bal}$ to be equal to zero, so we need to subtract a term equal to $S^\mathbf{raw}$, which is precisely $\left(R \star U \right) \cdot \Sigma D / \Sigma U$.

\subsection{Variance of the balanced sky image}\label{sec:variance}

Another useful quantity is represented by the variance of the balanced sky image $\text{Var}(S^\mathrm{bal}_{ij})$, which allows for the calculation of the signal-to-noise ratio (SNR) image and thus for the computation of the significance of the signal observed by each sky pixel. 

Let:
\begin{itemize}[leftmargin=0.1in]
    \item[] $\Lambda$ the array of the expected values of $D$
    \item[] $S = (R \star D)$  the unbalanced sky image
    \item[] $\xi = (R \star U)$ the balancing array 
    \item[] $\beta = \sum U$ the total active detector elements
    \item[] $\delta = \sum D$ the total detector counts
\end{itemize}
then each element of $S^\mathrm{bal}$ can be written as

\begin{equation}
S^\mathrm{bal}_{ij} = S_{ij} - \xi_{ij} \cdot \frac{\delta}{\beta}
\end{equation}

Since $S^\mathrm{bal}_{ij}$ is a linear combination of Poisson variables, and $\delta$ is their sum, using error propagation we can write
\begin{equation}
\text{Var}(S^\mathrm{bal}_{ij}) = \text{Var}(S_{ij}) + \left( \frac{\xi_{ij}}{\beta} \right)^2 \text{Var}(\delta) - 2 \cdot \frac{\xi_{ij}}{\beta} \cdot \text{Cov}(S_{ij}, \delta)
\end{equation}
where:
\begin{align*}
\text{Var}(S_{ij}) &= \sum_{k,l} R_{kl}^2 \Lambda_{i+k,j+l} \\
\text{Var}(\delta) &= \sum_{k,l} \Lambda_{kl} \\
\text{Cov}(S_{ij}, \delta) &= \sum_{k,l} R_{kl} \Lambda_{i+k,j+l}
\end{align*}

Therefore, expressing $\text{Var}(S^\mathrm{bal}_{ij})$ in the cross-correlation notation:

\begin{equation}\label{eq:variance}
\text{Var}(S^\mathrm{bal}_{ij}) = (R^2 \star \Lambda)_{ij} + \left( \frac{(R \star U)_{ij}}{\sum U} \right)^2 \cdot \sum \Lambda - 2 \cdot \frac{(R \star U)_{ij}}{\sum U} \cdot (R \star \Lambda)_{ij}
\end{equation}
where $R^2$ is the element-wise square of $R$.

As stated above, $S^\mathrm{bal}$ and $\text{Var}(S^\mathrm{bal})$ can be used to compute the SNR image of the observed field. An example of such a SNR image is shown in Section~\ref{sec:iros} (Figure~\ref{fig:galactic_center}).

\subsection{Mask code and design}\label{code}

CAI instruments are often based on bi-dimensional imaging detectors with squared or hexagonal pixels (or spatial resolution elements).
Perfect mask codes, characterized by a cyclic autocorrelation function approximating a delta function, have been implemented, for example, using cyclic difference sets \cite{Baumert1971}. Construction methods to obtain 2-dimensional rectangular, squared or hexagonal arrays have been implemented already in the late 1970s (see reviews \cite{Caroli1987,Accorsi2001,Goldwurm2022} and references therein).

Uni-dimensional coded aperture systems have also been employed in space-borne astrophysics experiments, like the All-Sky Monitor (ASM) on-board RXTE \cite{Levine1996} and SuperAGILE on-board the AGILE satellite \cite{Feroci2007,Feroci2010}. SuperAGILE exploited two pairs of orthogonally oriented uni-dimensional cameras, which combined linear (1-d) coded masks with silicon microstrip detectors.
To enhance imaging capabilities, each camera pair implemented a mask-antimask configuration (see, e.g., \cite{Braga2020}), with the mask code generated using a cyclic difference set of 787 elements.

Coded masks have also been built with elongated (i.e., rectangular) mask elements, such as the SL~1501 experiment \cite{Proctor1979} or the eXTP/WFM. The latter, derived from the LOFT design \cite{Brandt2014}, also employs a code based on cyclic difference set, built in this case from biquadratic residues (theorem 5.16 in \cite{Baumert1971}). The code is composed by 16640 elements and folded \cite{Caroli1987,intZand1994, Busboom1998} in 16 rows with height of 16.4~mm, each one containing 1040 code elements with a pitch of 0.25~mm \cite{Hernanz2018}. The choice of the mask element pitch is based on the detector spatial resolution \cite{Evangelista2014, Zwart2022} and results from the optimization of angular resolution, positioning accuracy and sensitivity as described in \cite{Skinner2008}.
It is worth noticing that, although the original eXTP/WFM mask open fraction (i.e., the ratio between the number of open elements and the number of total mask elements) is 0.25, manufacturing and thermo-mechanical requirements led to the insertion of 15 solid ribs of 2.4~mm width. Such ribs reduce the effective mask open fraction from 0.25 to $\sim$0.21.

As it will be described in Section~\ref{sec:design}, similarly to eXTP/WFM, the LEM-X mask is also characterized by the presence of mechanical ribs. 
The presence in the mask code of such solid, opaque ribs, requires a modification of the decoding array $R$ \cite{FenimoreCannon1978,Shutler2013}. In particular, in the following, we will consider a decoding array $R$ constructed as follows:

\begin{equation}\label{eq:r_ij}
    R_{ij} = 
    \begin{cases}
        1 & \text{if $M_{ij}$ is an open mask element} \\
        -\left( 1 - \frac{1}{N \cdot f_{o}} \right) \frac{f_{o}}{1-f_{o}} &  \text{if $M_{ij}$ is a closed mask element}\\
        0 &   \text{if $M_{ij}$ corresponds to a rib element}\\
    \end{cases}
\end{equation}
where $N$ is the total number of mask elements and $f_{o}$ is the original mask open fraction (i.e., the open fraction of the mask without the ribs). This choice of values for $R_{ij}$ ensures that the reconstructed image conserves the flux, has no pedestal around the sources, and it is locally unbiased \cite{SkinnerPonman1994}.

\section{LEM-X mask design}\label{mask_design}
\subsection{Flow-down of the LEM-X camera-level requirements}
In Table~\ref{camera_reqs} we report a subset of the main camera-level requirements, as flowed-down from the LEM-X scientific requirements summarized in Table~\ref{tab:sci_reqs}. These camera-level requirements drive the instrument design and consequently the design of the coded mask subsystem.

\begin{figure}[h!]
\centering
\includegraphics[width=0.92\textwidth]{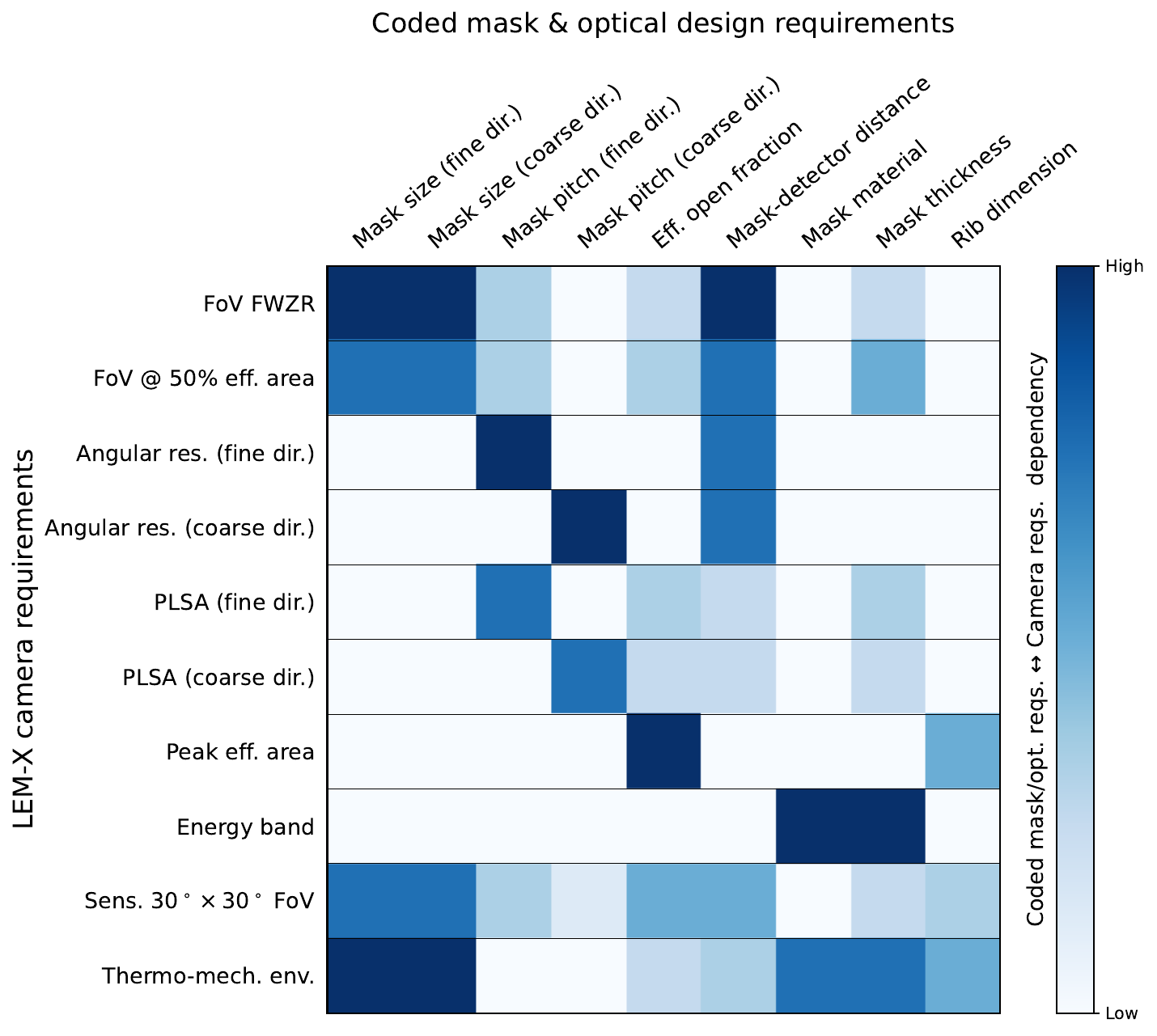}
\caption{Dependency matrix of the coded mask and optical design requirements in Table~\ref{mask_reqs} and the higher level camera requirements in Table~\ref{camera_reqs}.}\label{req_matrix}
\end{figure}

It is worth noticing that almost all the camera requirements in Table~\ref{camera_reqs} impact the definition of different coded mask requirements. For such a reason, the final coded mask design is the result of a trade-off activity, performed taking into account the different contributions to the final performance.
Although a detailed discussion of this trade-off activity is beyond the scope of this work, we present in Figure~\ref{req_matrix} a visual representation of the dependency matrix between the coded mask and the camera requirements.

When the LEM-X detector plane design and SDD performance described in Section~\ref{sec:instrument_design} are taken into account (e.g., arrangement of the detectors, DA sensitive area,  asymmetric SDD spatial resolution), the higher-level requirements in Table~\ref{camera_reqs} can be flowed-down to the coded mask and optical design requirements reported in Table~\ref{mask_reqs}.

\begin{table}[h!]
\caption{LEM-X camera requirements}\label{camera_reqs}%
\begin{tabular*}{\textwidth}{@{\extracolsep\fill}llll@{}}
\toprule
Requirement & Value  & Note\\
\midrule
\multirow{2}{*}{Field of View}  &$90^\circ\times90^\circ$     &  full width at zero response (FWZR)\\
                                &$\geq 40^\circ\times40^\circ$     &  at 50\% effective area\\
\multirow{2}{*}{Angular resolution} &$\leq$ 5$'$ FWHM            &  fine resolution direction\\
                                    &$\leq$ 5$^\circ$ FWHM            &  coarse resolution direction\\
\multirow{2}{*}{Point source location accuracy} &1$'$ at 90\% conf. level & fine resolution direction\\
                                                &1$^\circ$ at 90\% conf. level & coarse resolution direction\\
Peak effective area                  & $\geq$ 65 cm$^2$   &  at 8~keV \\
Energy band                     & 2--50 keV   &    \\
Sensitivity                     & $\leq$ 1000 mCrab   & 5$\sigma$, 1~s exposure, $30^\circ \times 30^\circ$ FoV\\
Operative temperature range    & $-$50~$^\circ$C $\leq$ T $\leq$ $-$20~$^\circ$C \\
Quasi-static loads              & 25 g & per axis \\
Sine loads                      & 25 g & per axis, up to 110 Hz\\
Minimum natural frequency       & $\geq$120~Hz & \\
Random loads                      & 14.1 grms & TBC\\
\botrule
\end{tabular*}

\end{table}

\begin{table}[h!]
    \caption{LEM-X coded mask and optical design requirements}\label{mask_reqs}%
    \begin{tabular*}{\textwidth}{@{\extracolsep\fill}llll@{}}
        \toprule
        Requirement & Value  & Note\\
        \midrule
        Mask size                       & $\sim260\times260$ mm$^2$ & \\
        \multirow{2}{*}{Mask pitch}     & $p_x$ = 250 $\mathrm{\upmu}$m & fine resolution direction\\
                                        & $p_y = 15.5$ mm & coarse resolution direction  \\
        Mask effective open fraction    & $0.4 \leq f_e \leq 0.5$& \\
        Mask-detector distance          & 202.975 mm& top of detector plane-mid mask plane\\
        Mask material                   & Tungsten   &    \\
        Mask thickness                  & 150 $\upmu$m &    \\
        Solid rib height                & 2.5 mm &  \\
        \botrule
    \end{tabular*}
\end{table}

\subsection{Mask code}\label{sec:design}
The requirements in Table~\ref{camera_reqs} result in a mask size of $\sim260\times260$~mm$^2$. Considering the 250~$\upmu$m mask pitch in the fine direction ($p_x$) and the 15.5~mm pitch in the coarse direction ($p_y$), this implies a total of $1040\times17 = 17680$ mask elements, for an overall dimension of $260.0\times263.5$~mm$^2$. The closest next prime number ($L$) to 17680 is $L=17681$, which satisfies Eq. 10 in \cite{GottesmanFenimore1989} for the construction of a Modified Uniformly Redundant Array (MURA) with $m=4420$.

The uni-dimensional binary sequence $A$ for building the mask code is defined as:
\begin{equation}\label{eq:a_k}
    A_{k} = 
    \begin{cases}
        0 &  \text{if $k = 0$ } \\
        1 & \text{if $k=(l^2~\mathrm{mod}~L)$ for $0\leq l < L$} \\
        0 &   \text{otherwise}\\
    \end{cases}
\end{equation}
where 1 represents the transparent (open) elements, and 0 the opaque (closed) ones. 
The resulting open fraction of $f_o = 0.5$ represents a good compromise for spectral-timing studies of transient events and the monitoring of compact galactic sources \cite{intZand1994, Skinner2008}.

The $1040\times17$ mask can be obtained by taking the first 17680 elements of the binary MURA code $A$ and folding them in 17 rows, each one of 1040 elements (columns). 

The use of a MURA (instead of an URA-based mask) leads to a further condition to be added to Eq.~\ref{eq:r_ij}. In particular, the first element of the decoding array $R$ must be forced to 0 \cite{GottesmanFenimore1989}, which leads to: 
\begin{equation}\label{eq:r_ij_mura}
    R_{ij} = 
    \begin{cases}
        0 &  \text{if $i,j = 0$ } \\
        1 & \text{if $M_{ij}$ is an open mask element} \\
        -\left( 1 - \frac{1}{N \cdot f_{o}} \right) \frac{f_{o}}{1-f_{o}} &  \text{if $M_{ij}$ is a closed mask element and $i,j \neq 0$}\\
        0 &   \text{if $M_{ij}$ corresponds to a rib element}\\
    \end{cases}
\end{equation}

Because MURA codes have the useful property of being symmetric with respect to their first element \cite{GottesmanFenimore1989}, we decided to apply a cyclic shift to the code before folding it, this in order to achieve a more symmetric and homogeneous mask response. Such a symmetry is beneficial in ensuring a uniform and consistent response at the Unit level and at the Instrument level.
\begin{figure*}[h!]
\centering
\subfloat[]{\includegraphics[width=\textwidth]{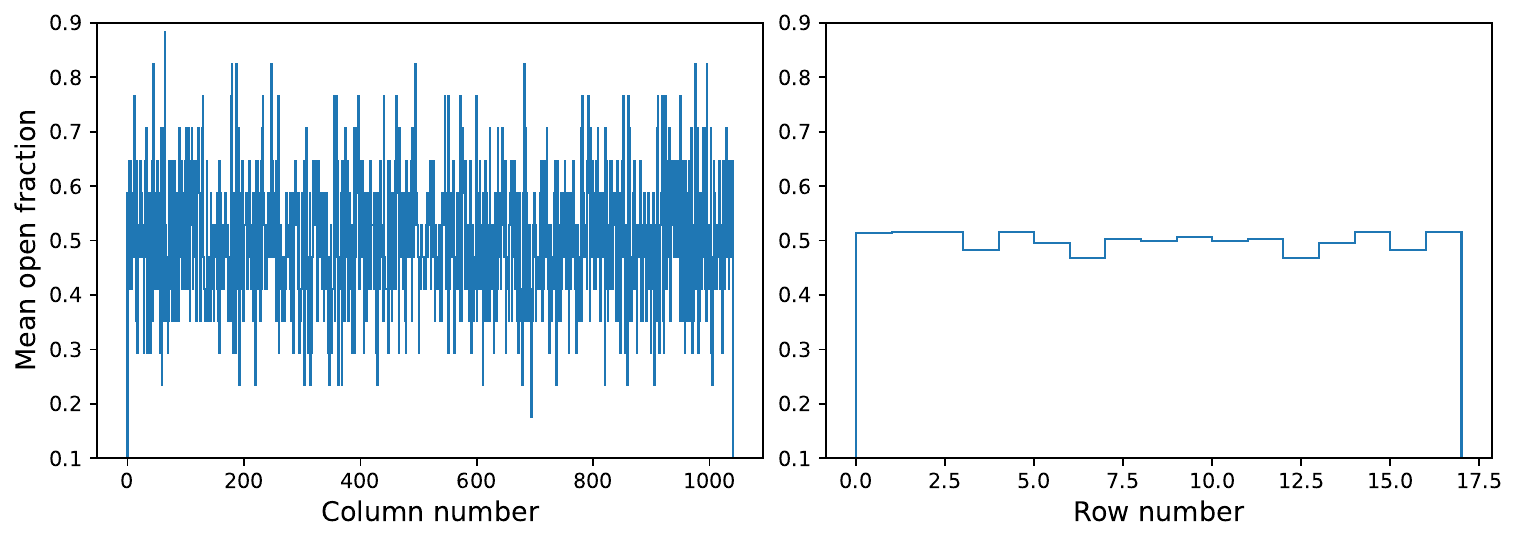}}\\
\subfloat[]{\includegraphics[width=\textwidth]{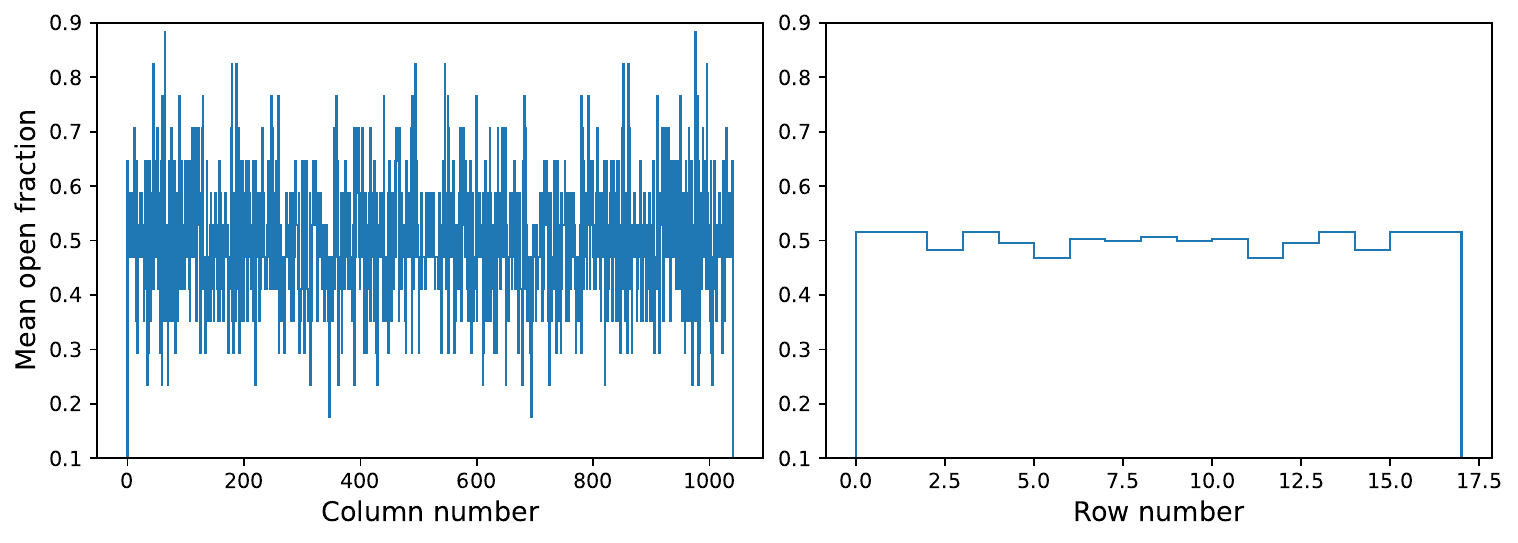}}
\caption{Mean mask open fraction $\overline{f_o}$, averaged along the mask columns (left) and rows (right), before (a) and after (b) applying a cyclic shift to the MURA code to exploit its symmetry with respect to the first code element. See text for more details.}
\label{fig:homogeneity}
\end{figure*}

A proxy of the mask response homogeneity can be obtained by estimating the mean mask open fraction $\overline{f_o}$ of each row and column in the folded code. The result of this operation, performed for the 1040 mask columns and the 17 mask rows before and after the cyclic shift, is shown in Figure~\ref{fig:homogeneity}.

It is important to note that the condition for $i,j=0$ presented in Eq.~\ref{eq:r_ij_mura} shall be applied prior to the application of any code shift. Consequently, the decoding matrix $R$ must be constructed from the original (folded) MURA code and then shifted in the same manner as the mask.

The introduction of the 2.5~mm solid ribs results, as expected, in a reduction of the effective open fraction ($f_e$) with respect to the original one. When considering the effective mask dimensions after the introduction of the ribs (calculated as the maximum distance between open elements), the LEM-X mask has dimensions of $260\times261$~cm$^2$ (fine  $\times$ coarse resolution directions) and $f_e = 0.42$.

In Figure~\ref{fig:LEMX_mask} we report the mask with the solid, opaque ribs highlighted in light blue.

\subsection{System performance}

\begin{figure*}[h!]
\centering
\subfloat[]{\includegraphics[width=0.6\textwidth]{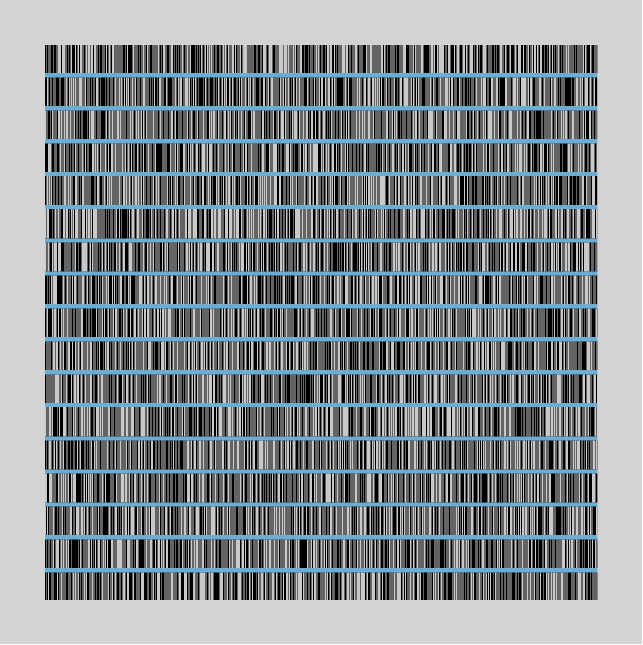}\label{fig:LEMX_mask}}\\
\subfloat[]{\includegraphics[width=0.65\textwidth]{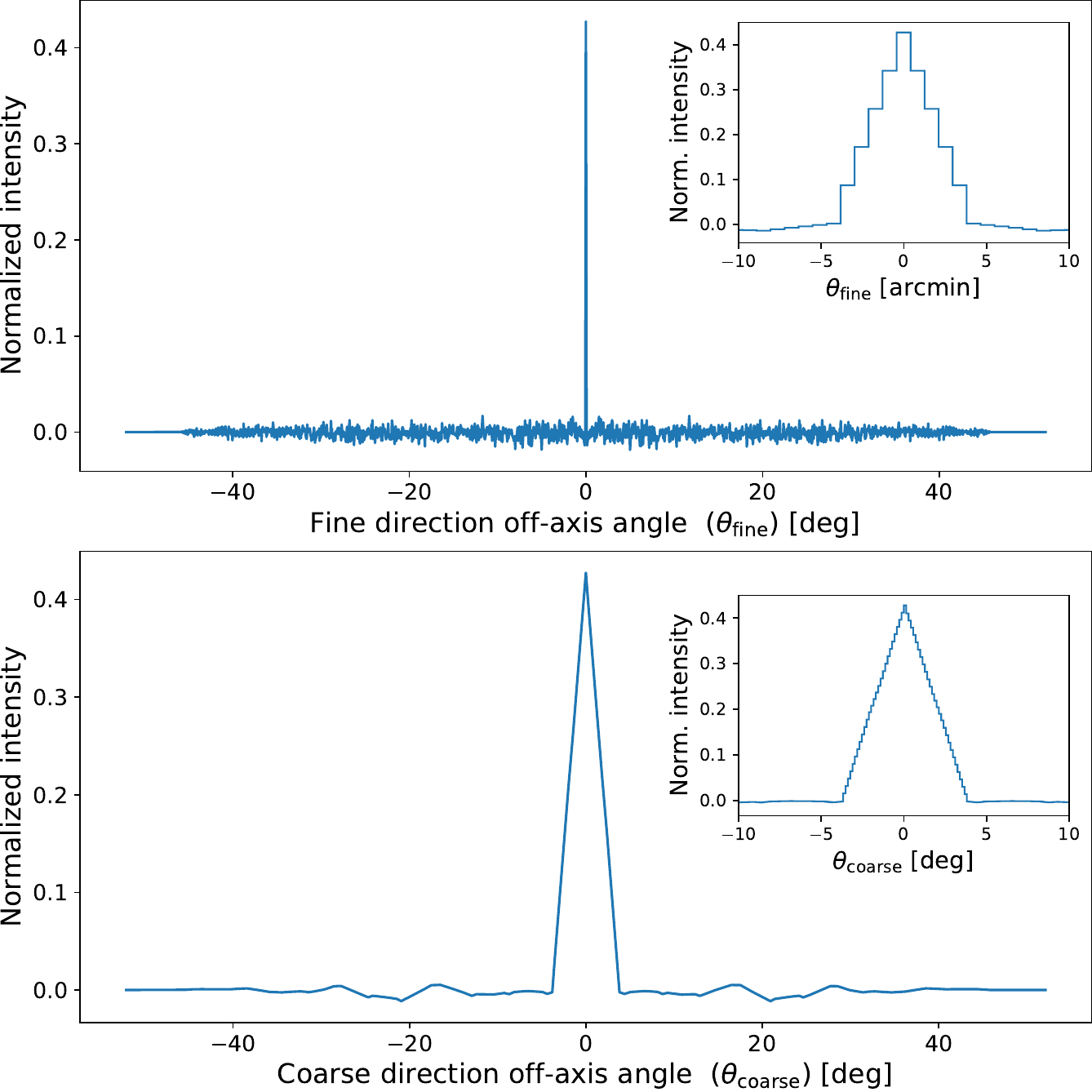}\label{fig:psf}}
\caption{(a) LEM-X coded mask design. Open mask elements are represented in black, while closed mask elements and ribs are shown in gray and light blue respectively.\\ (b) Uni-dimensional camera SPSF (fine and coarse resolution directions) obtained using Eq.~\ref{eq:deconvolution}. See text for more details.}
\end{figure*}

By using Eq.~\ref{eq:deconvolution}, and replacing the matrix $D$ with a matrix $D\mathrm{'}_{ij} = M_{ij}\, \odot \, U_{ij}$ representing the mask array ($M$) weighted for the detector efficiency matrix $U$, it is possible to obtain the system point spread function (SPSF), which represents how the system reconstructs a single on-axis point source \cite{FenimoreCannon1978}. The LEM-X camera SPSF is shown in Figure~\ref{fig:psf}. For the sake of clarity, the SPSF is shown in two one-dimensional image slices, obtained in the center of the FoV along the fine resolution (top panel) and coarse resolution (bottom panel) directions. The SPSF shows a single peak, with angular dimensions in fine and coarse direction of $\sim$4.4~arcmin and 4.8$^\circ$ respectively, and with a normalized intensity matching the mask effective open fraction. It can be seen that the SPSF is characterized by almost flat sidelobes, thus indicating a low inherent noise (coding noise \cite{intZand1992, Goldwurm2022}).
The presence of such a coding noise in the SPSF is mainly due to the design of the LEM-X camera, characterized by a detector plane $\sim$1.65 times smaller than the mask, and a mask code which is not the cyclic repetition of a basic (smaller) pattern (see \cite{Caroli1987} and references therein). These design choices are the result of a trade-off activity carried out taking into account the large FoV of each camera and its sensitivity. The two characteristics favor the use of a single basic mask pattern, which avoids the presence of ghost peaks \cite{Goldwurm2022, Skinner2008} in the reconstructed images and maximizes the system signal-to-noise ratio, especially in the case of crowded fields and/or in presence of particularly intense sources.

\begin{figure*}[h!]
\centering
\subfloat[]{\includegraphics[width=0.75\textwidth]{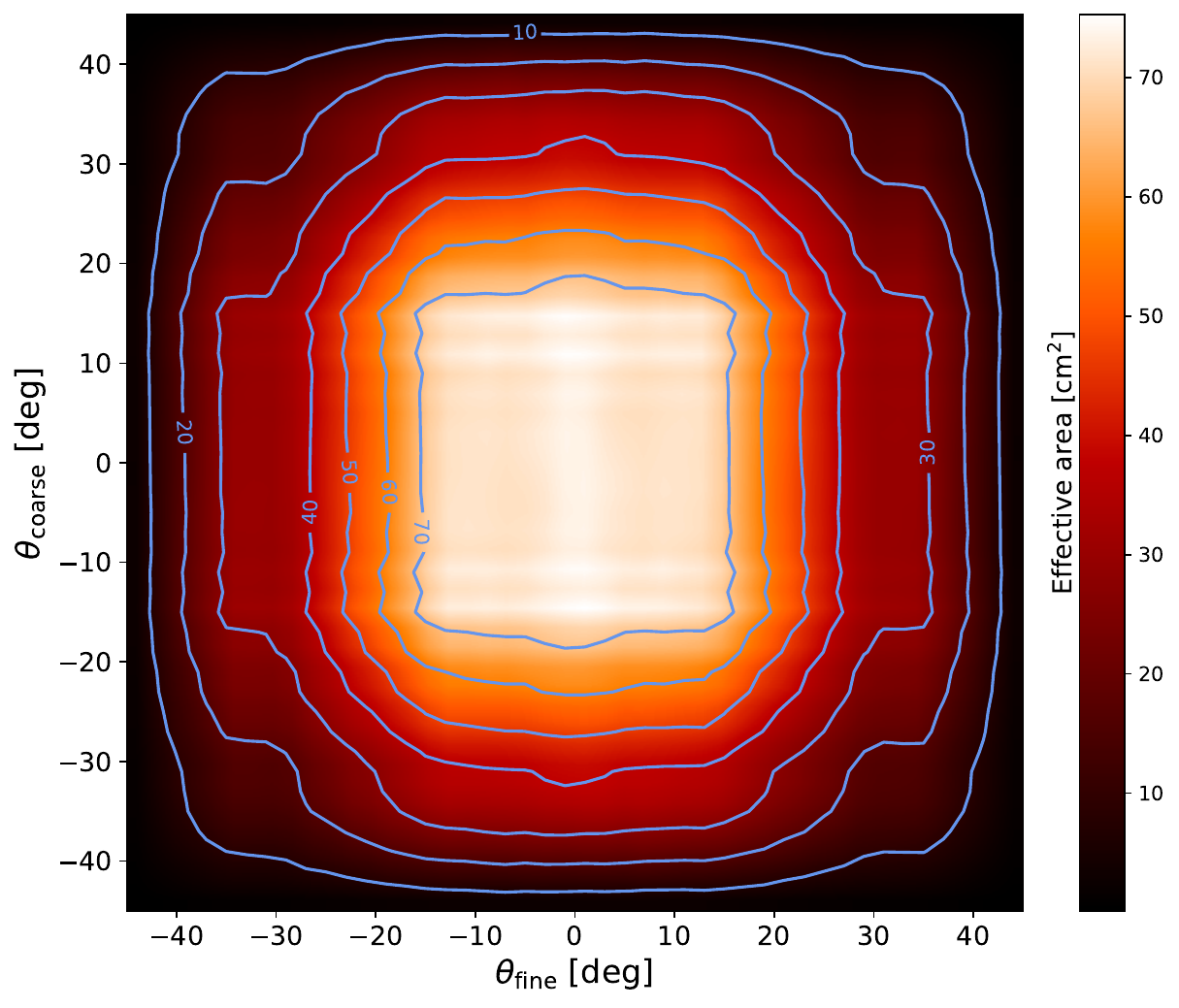}\label{fig:eff_area}}\\
\subfloat[]{\includegraphics[width=0.75\textwidth]{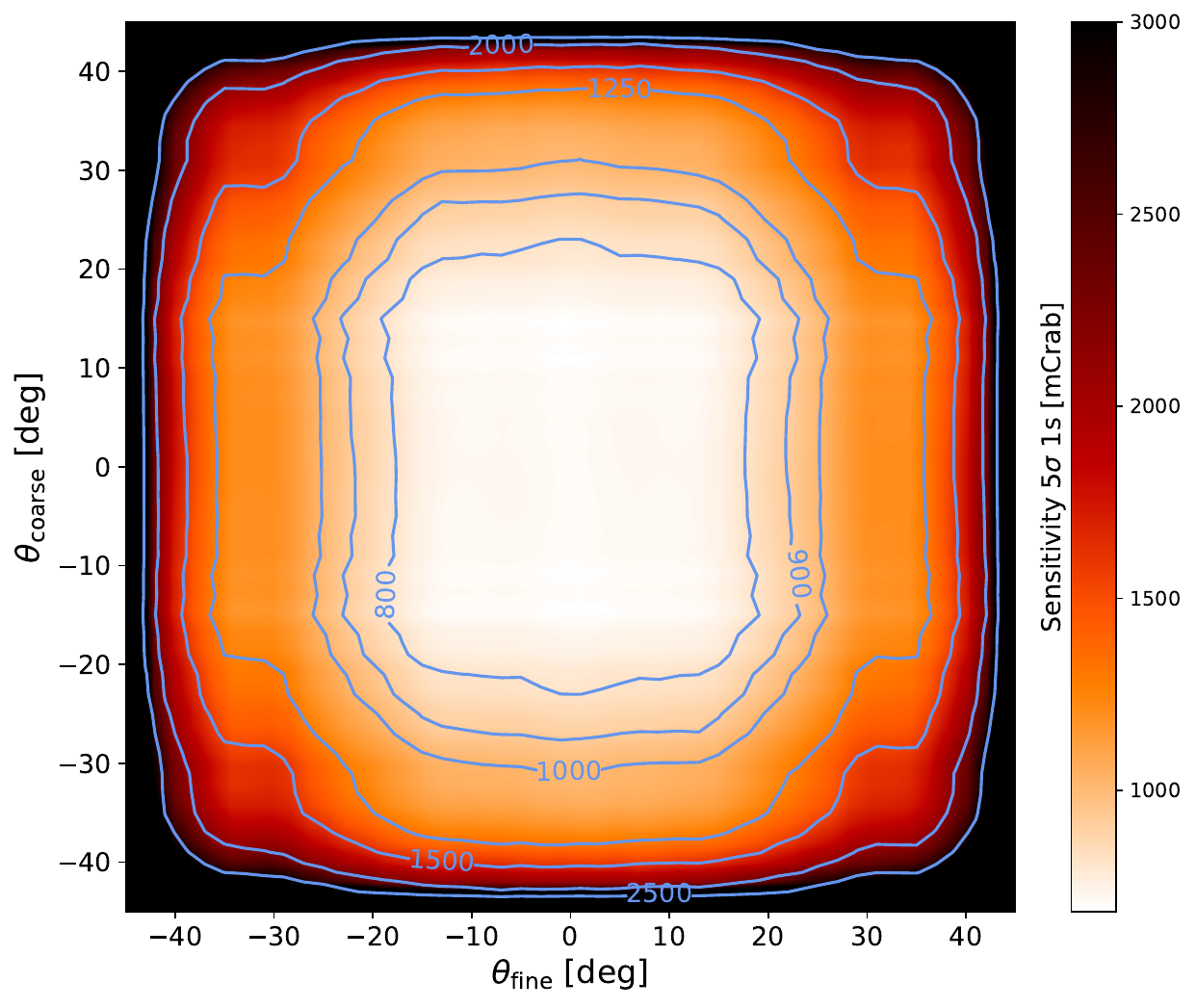}\label{fig:sens}}
\caption{(a) Camera effective area as a function of the off-axis angle in the fine and coarse coding directions. (b) Camera sensitivity (for an isolated source in presence of Cosmic X-ray Background) as a function of the off-axis angle in the fine and coarse coding directions.}
\label{fig:eff_area_sens}
\end{figure*}

The resulting camera effective area and sensitivity maps, as a function of the off-axis angle, are shown in Figure~\ref{fig:eff_area_sens}. As it can be seen from Figure~\ref{fig:eff_area}, the peak effective area is $>72~ \mathrm{cm}^2$, with a FoV at 50\% effective area of $\sim 44^\circ \times 44^\circ$ and of $\sim 90^\circ \times 90^\circ$ at zero response. Such performance guarantees satisfactory margins (12\% and 20\%, respectively) with respect to the camera requirements listed in Table~\ref{camera_reqs}.

The sensitivity map shown in Figure~\ref{fig:sens} has been analytically calculated using Eq.~\ref{eq:snr} below, derived from Eq.~13 in \cite{Skinner2008}:

\begin{equation}\label{eq:snr}
    \mathrm{SNR}(\theta_\mathrm{f}, \theta_\mathrm{c}) = \Delta * \frac{C_\mathbf{S}}{ f(\theta_\mathrm{f}, \theta_\mathrm{c})} \cdot 
    \displaystyle \sqrt{ 
        \frac{
            \displaystyle A_\mathrm{eff}(\theta_\mathrm{f}, \theta_\mathrm{c}) \cdot \left[ 1 - f(\theta_\mathrm{f}, \theta_\mathrm{c}) \right] 
        }
        {
            f(\theta_\mathrm{f}, \theta_\mathrm{c}) \cdot \displaystyle \frac{C_\mathbf{S}+C_\mathbf{B}}{f_\mathrm{e}}
        }
    }
\end{equation}
where $C_\mathbf{S}$ and $C_\mathbf{B}$ are the source and background counts per unit area, $f_e$ is the mask open fraction, $A_\mathrm{eff}(\theta_\mathrm{f}, \theta_\mathrm{c})$ and $f(\theta_\mathrm{f}, \theta_\mathrm{c})$ are the exposed effective area and open fraction for a source located at ($\theta_\mathrm{f}, \theta_\mathrm{c}$) off-axis angles, for the fine and coarse resolution directions respectively. Cosmic X-ray Background (CXB) and source count-rates have been estimated by means of the \emph{WISEMAN} simulator \cite{Ceraudo2024wiseman}.
It is worth noticing that $A_\mathrm{eff}(\theta_\mathrm{f}, \theta_\mathrm{c})$ takes into account the off-axis dependent vignetting introduced by the mask finite thickness, as well as the effective detector area illuminated by an off-axis source. On the other hand, $f(\theta_\mathrm{f}, \theta_\mathrm{c})$ takes into account the open fraction of the mask portion projected onto the detector at a specific off-axis angle, again considering the thickness-induced vignetting.

Lastly, let's analyze the factor $\Delta$, which is the \emph{coding power} introduced in Section~6 of \cite{Skinner2008}. This factor takes into account the loss in sensitivity caused by the finite detector resolution, as compared with the ideal case where the detector is characterized by a perfect (i.e. infinite) spatial resolution.
In practice, $\Delta$ can be calculated numerically by comparing the SPSF obtained with a perfect detector resolution with the SPSF obtained by ``blurring'' (i.e. convolving) the detector image with a kernel representing the actual spatial resolution. For the LEM-X camera this calculation results in a coding power $\Delta \simeq 0.85$.

The \emph{WISEMAN} photon-by-photon simulator was also employed to benchmark the analytical sensitivity calculation described above at various positions within the FoV. 
This comparison aimed to quantify the impact of the simplifications in Eq.~\ref{eq:snr}, which neglects energy-dependent effects such as mask transparency, detector spatial resolution, and photon absorption depth in the silicon bulk.
The analysis highlighted a very good agreement between the two methods, with the analytical calculation in Eq.~\ref{eq:snr} overestimating by less than 5\% the sensitivity computed using Eq.~\ref{eq:deconvolution} and Eq~\ref{eq:variance} with photon-by-photon simulated data. 

The LEM-X camera peak sensitivity for an isolated source, obtained on-axis, is better than 700~mCrab in 1~s at 5$\sigma$ (corresponding to $\sim$3.2~mCrab in 50~ks), while the required sensitivity (better than 1000~mCrab in 1~s at $5\sigma$) is obtained in a $\sim 45^\circ \times 45^\circ$ FoV. 

\clearpage

\section{Mechanical design and simulations}\label{thermomech}
\begin{figure*}[ht!]
\centering
\subfloat[]{\includegraphics[width=0.36\textwidth]{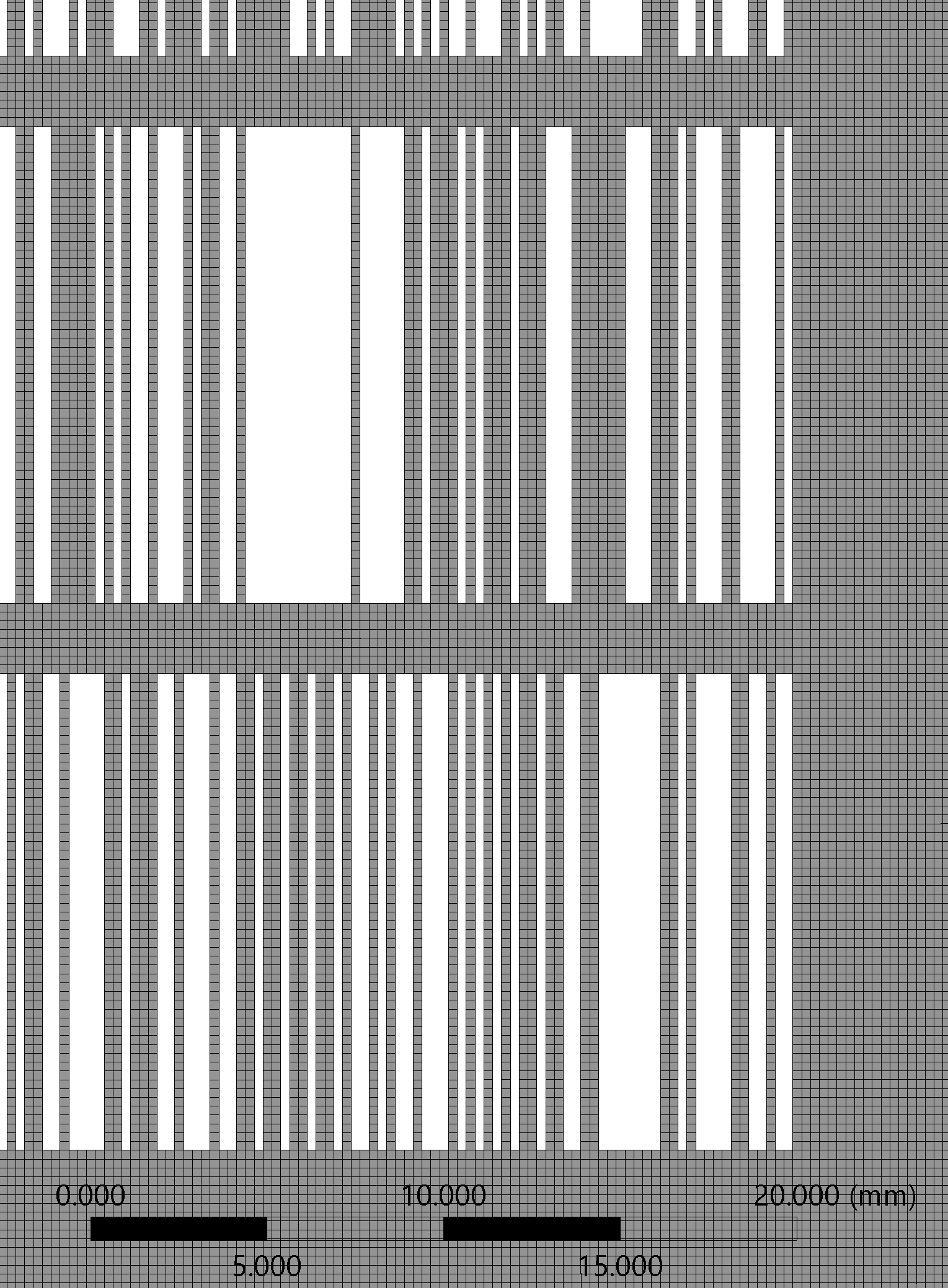}\label{fig:fem_mesh}}
\hspace{0.02\textwidth}
\subfloat[]{\includegraphics[width=0.6\textwidth]{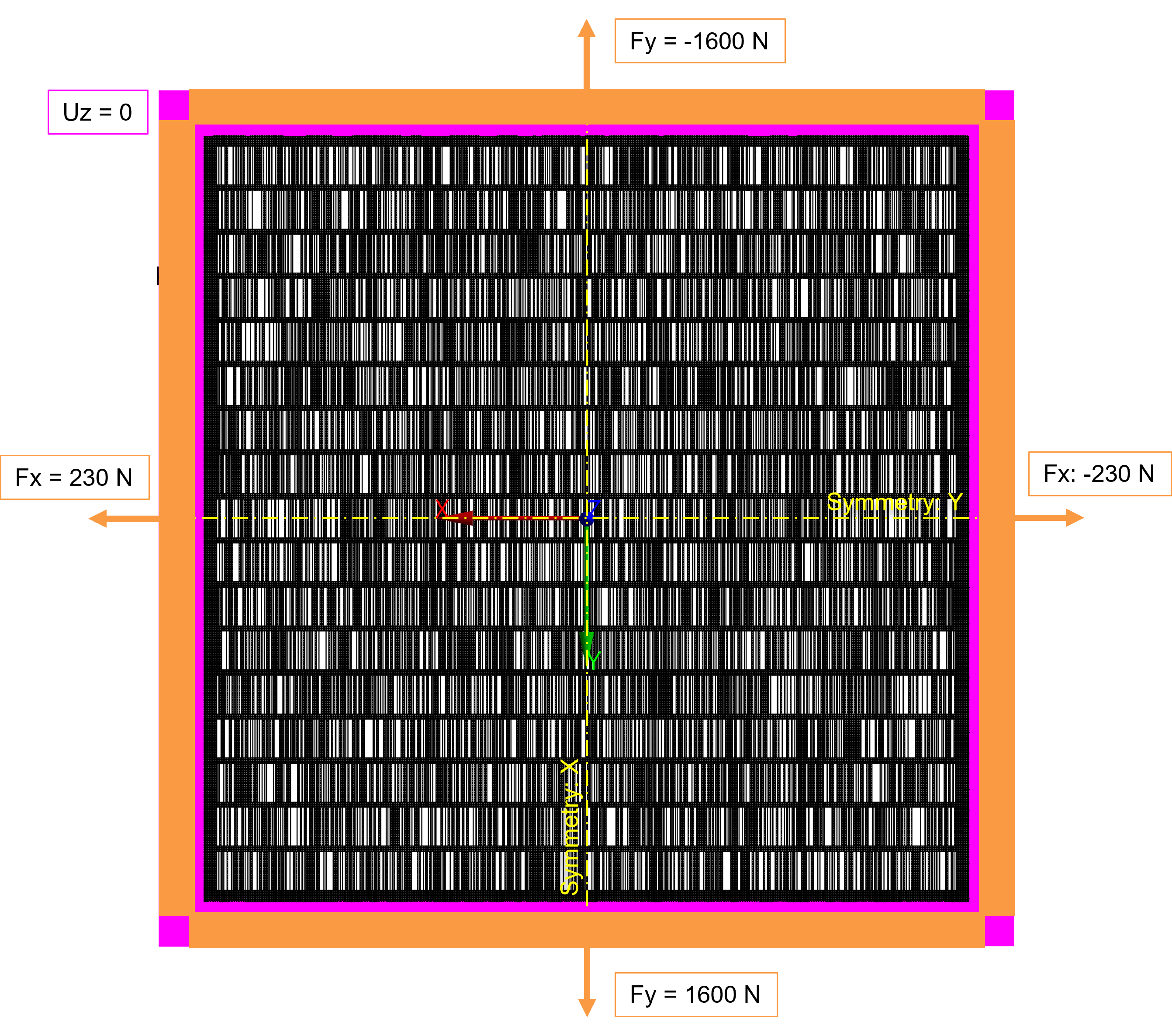}\label{fig:fem_model}}
\caption{Pretensioned mask FEM model: (a) mesh detail, (b) loads and boundary conditions.}
\label{fig:fem}
\end{figure*}
The mechanical design of the LEM-X coded mask is driven by the need to guarantee structural integrity and dimensional stability throughout both launch and operational environments. The design trade-off led to the selection of a 150~µm thick tungsten sheet as reported in Table~\ref{mask_reqs}. To avoid the introduction of additional material in the optical path, which would degrade performance, the mask is supported by applying a controlled pretension to the tungsten sheet itself. Finite Element Method (FEM) simulations were performed to validate this concept.

The finite element model was developed in ANSYS Mechanical 2022R2. The mask was modeled using shell (2D) elements with a characteristic size of 0.25~mm, as shown in Figure~\ref{fig:fem_mesh}. The applied loads and boundary conditions (Figure~\ref{fig:fem_model}) are symmetric with respect to the two vertical planes. Although the slits slightly break the symmetry in the mask geometry, their distribution is sufficiently uniform to allow the structure to be treated as globally symmetric. For this reason, symmetry constraints were applied on the two planes to make the system statically determined. The full model was nevertheless simulated to capture possible local effects associated with the specific slit pattern. A comparative simulation without symmetry constraints, using weak spring stabilization, confirmed negligible differences in both displacement and stress fields.
\clearpage
\begin{figure*}[ht!]
\centering
\subfloat[]{\includegraphics[width=0.9\textwidth]{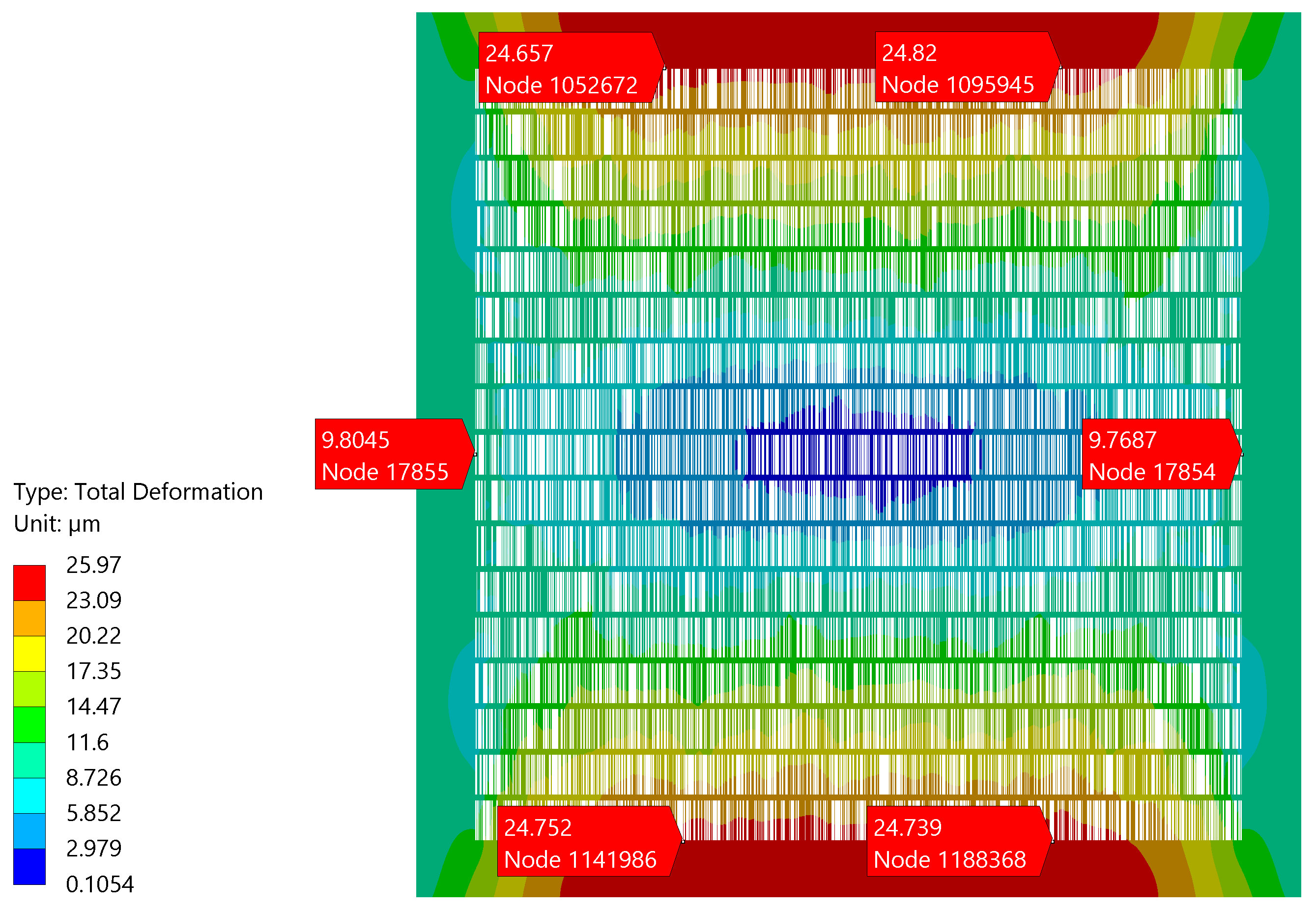}\label{fig:fem_displ}}
\\
\subfloat[]{\includegraphics[width=0.9\textwidth]{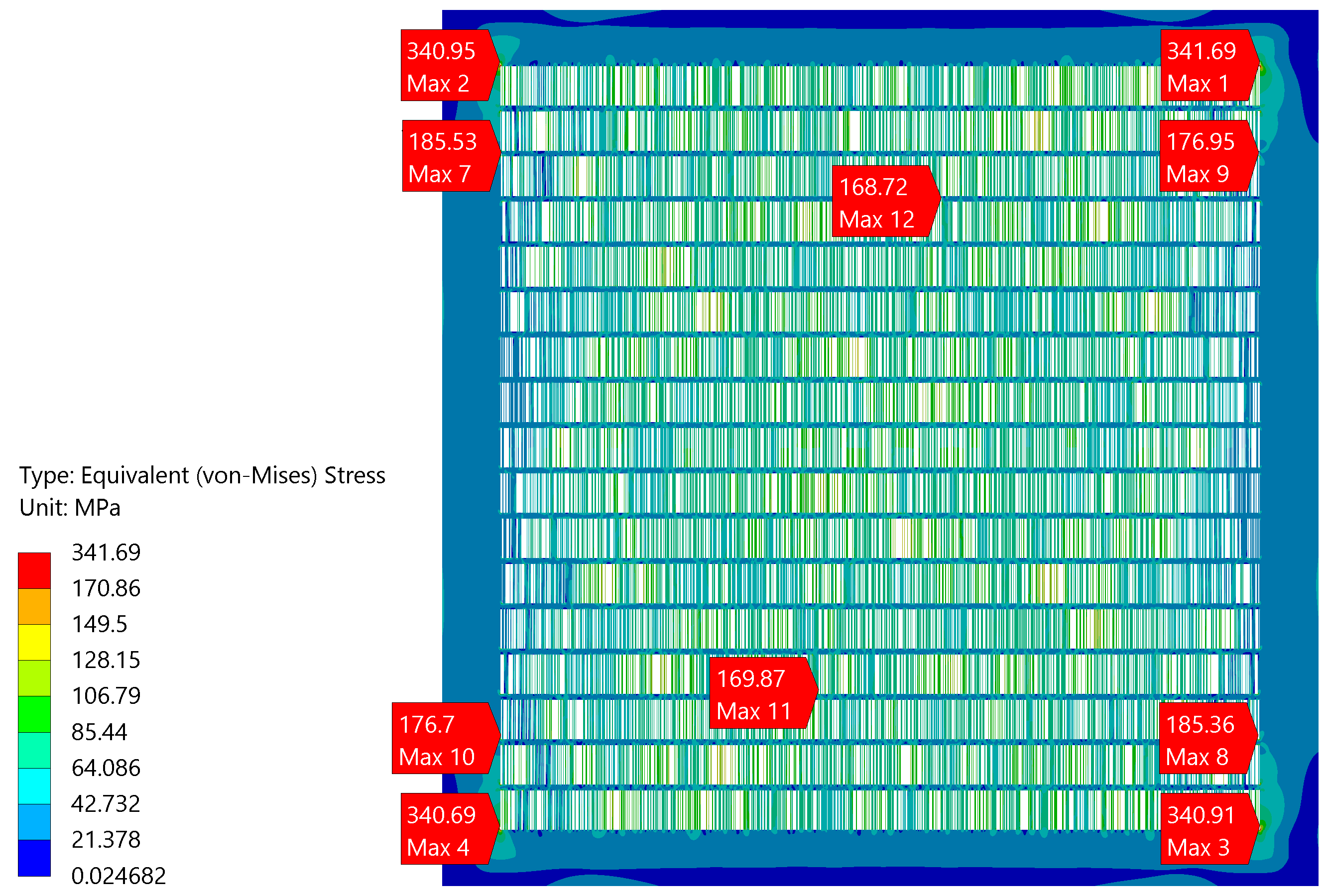}\label{fig:fem_stress}}
\caption{Pretensioned mask FEM results: (a) in-plane displacement magnitude, (b) equivalent stress distribution}
\label{fig:fem_results}
\end{figure*}
Four uniformly distributed forces were applied along the edges to pretension the mask (Figure~\ref{fig:fem_model}), with different values in the two orthogonal directions to account for the anisotropic stiffness of the structure. The pretension levels were defined through a trade-off between achieving a maximum in-plane displacement lower than 10~$\upmu$m in the fine direction and maintaining a first natural frequency above 120~Hz.

The FEM results for the static pretension load case are reported in Figure~\ref{fig:fem_results}. The analysis shows that the maximum in-plane displacements of the slits are below 10~$\upmu$m in the fine direction and 25~$\upmu$m in the coarse direction, with negligible impact on the optical performance. The corresponding stress distribution remains below 190~MPa over the full mask except for the corners, where the stress reaches 342~MPa. 

A random vibration analysis was then performed using the GEVS qualification spectrum, applying the acceleration load along the out-of-plane axis, which represents the worst-case scenario for the mask. The maximum equivalent stresses, calculated summing the 3-sigma random results to the static pretension values, do not exceed 412 MPa. 
This provides a significant margin against the yield strength for tungsten of about 750~MPa, demonstrating that the mask assembly can safely survive the expected launch vibration loads.

The response to a 1~g out-of-plane load was also simulated, resulting in a maximum out-of-plane displacement of 24~$\upmu$m, which does not pose concerns for performance during ground calibration.

The results confirm the feasibility of the pretensioned tungsten mask concept, demonstrating that the proposed configuration satisfies the required stiffness and strength criteria. A more detailed stress and thermoelastic analysis will be carried out once the actual pretensioning mechanism and support structure design has been consolidated, in order to validate the complete assembly under the expected thermo-mechanical environment.

\section{Imaging simulations}\label{sec:iros}
The imaging capabilities of both a single LEM-X camera and the two orthogonal cameras forming a Unit have been thoroughly investigated and validated using the \emph{WISEMAN} photon-by-photon Monte Carlo instrument simulator, alongside an Iterative Removal Of Sources (IROS) procedure \cite{Hammersley1992}. This procedure was specifically developed for the LEM-X project and is based on the imaging techniques described in Sections~\ref{sec:decoding} and~\ref{sec:variance}.

The LEM-X IROS pipeline \cite{Giancarli2025} draws upon the \emph{Bloodmoon}\footnote{\url{https://github.com/peppedilillo/bloodmoon}} and \emph{Darksun}\footnote{\url{https://github.com/EdoardoGiancarli/darksun}} Python packages. 
The \emph{Bloodmoon} package implements IROS at the Unit level (i.e., combining data from both cameras) and provides primitives for coded mask image reconstruction, source detection, and parameter optimization. These routines account for instrumental effects such as mask-induced vignetting and the detector spatial resolution in both fine and coarse directions. The \emph{Darksun} package focuses on image processing and the application of the IROS pipeline, including source validation against catalogs, statistical analysis of the images, and generation of scientific products such as count maps and SNR images at both camera and Unit levels.

\begin{figure}[h!]
\centering
\subfloat[]{\includegraphics[width=0.67\textwidth]{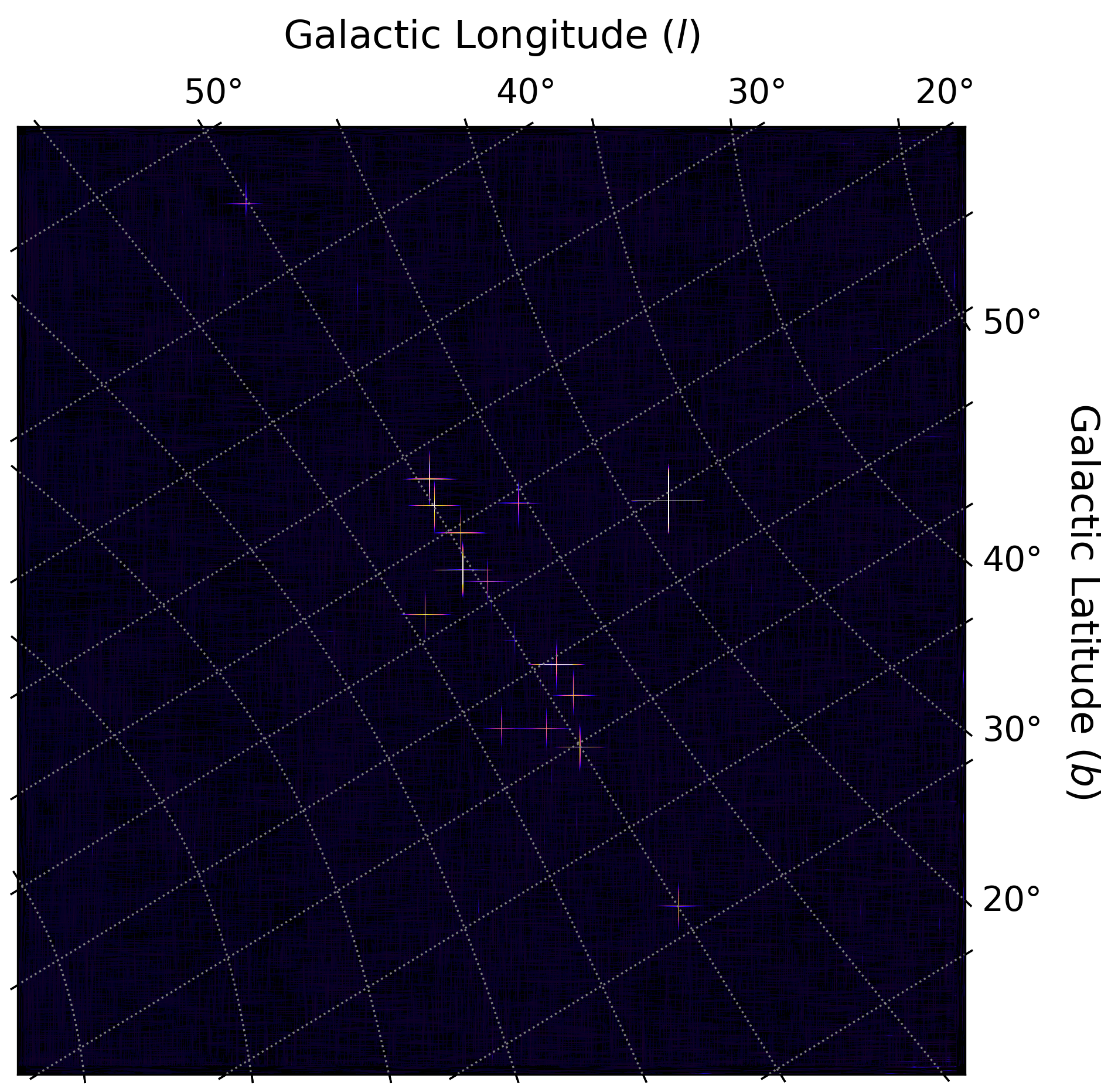}\label{fig:composed_sky}}\\
\subfloat[]{\includegraphics[width=0.67\textwidth]{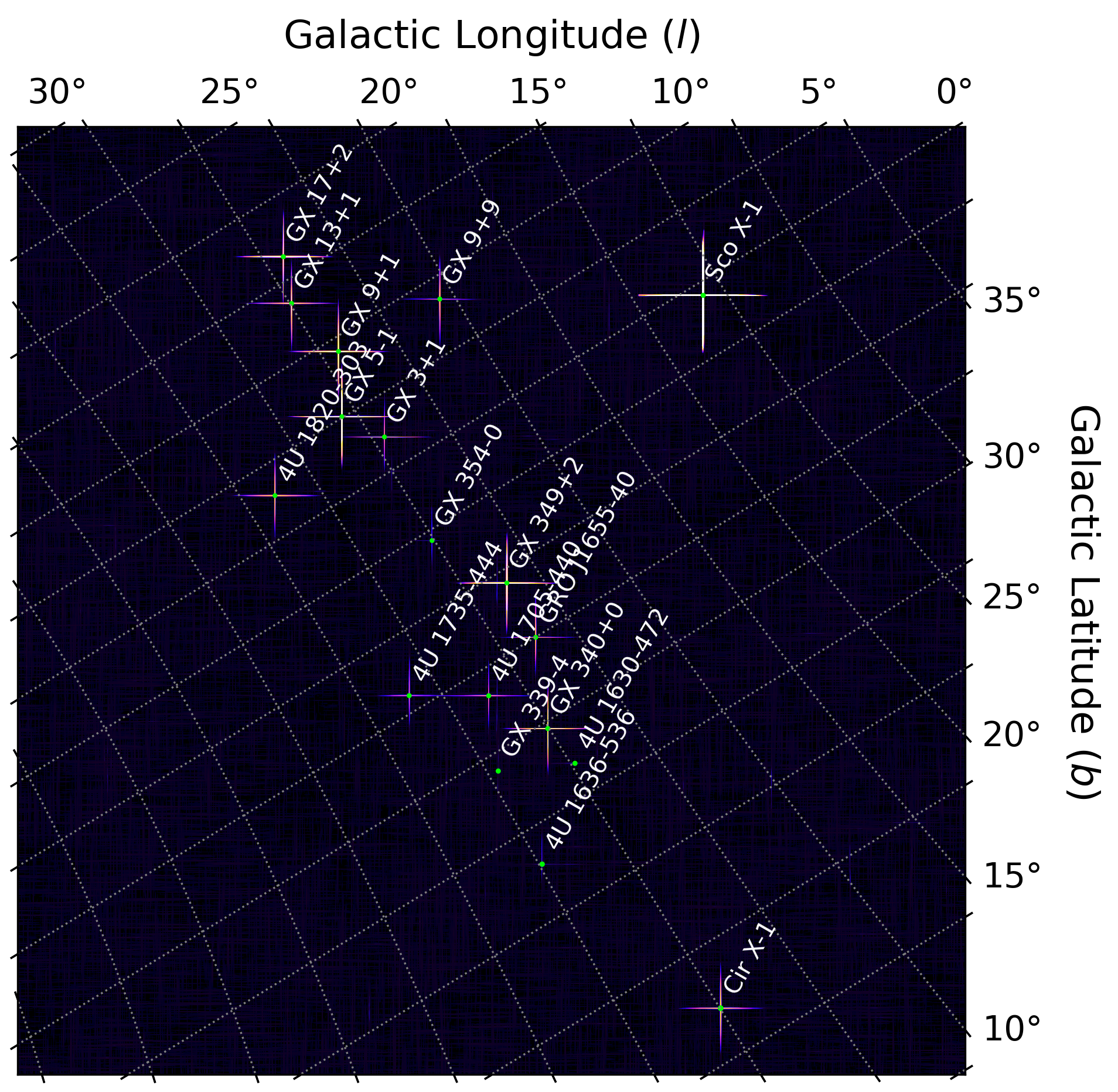}\label{fig:composed_sky_zoom}}
\caption{Simulation of 1~ks exposure on the galactic center, reconstructed with the \emph{Bloodmoon} and \emph{Darksun} packages. (a) Full FoV reconstructed SNR image obtained using Eq.~\ref{eq:deconvolution} and Eq~\ref{eq:variance}. (b) Zoom of the $\sim 45^\circ \times 45^\circ$ central FoV region with detected source names.}\label{fig:galactic_center}
\end{figure}

The results of a 1~ks observation of the Galactic Center, simulated using the RXTE/ASM source catalog\footnote{\url{https://heasarc.gsfc.nasa.gov/docs/xte/ASM/sources.html}} and the CXB spectral model from \cite{Ajello2008}, are summarized in Table~\ref{tab:iros_src_data}. A total of 20 sources with a two-cameras combined SNR~$\geq 5\upsigma$ have been detected and modeled by the IROS pipeline. The  reconstructed significance images, obtained by combining the SNR images simultaneously collected by the two orthogonal cameras in a Unit, is shown in Figure~\ref{fig:galactic_center}. 
The full field-of-view image (approximately $90^\circ \times 90^\circ$) is shown in Figure~\ref{fig:composed_sky}, while Figure~\ref{fig:composed_sky_zoom} provides a zoomed-in view of the Galactic Center region, highlighting the detected X-ray sources and demonstrating the instrument effective imaging performance and sensitivity.

\begin{table}[ht]
\caption{IROS detected sources}\label{tab:iros_src_data}
\tabcolsep=0.11cm
\begin{tabular*}{\textwidth}{@{\extracolsep\fill}lcccccccccc@{}}
\toprule
\multirow{2}{*}{Source name} & 
\multirow{2}{*}{Flux\footnotemark[1]}  & 
\multicolumn{3}{c}{Camera A} & 
\multicolumn{3}{c}{Camera B} &
\multirow{2}{*}{SNR\footnotemark[4]} & 
\multirow{2}{*}{$\Delta\theta$\footnotemark[5]} \\
 & & 
$\theta_\mathrm{f}\footnotemark[2]$ & $\theta_\mathrm{c}\footnotemark[2]$ & Counts\footnotemark[3] & 
$\theta_\mathrm{f}\footnotemark[2]$ & $\theta_\mathrm{c}\footnotemark[2]$ & Counts\footnotemark[3] & & \\
\midrule
Sco X-1        & 32.80 & 12.20 & 21.03 & 1055475 & 21.03 & 12.20 & 943406 & 993.6 & 0.03 \\
GX 5-1         & 2.52  & 3.81  & -3.53 & 114831  & -3.53 & 3.81  & 116247 & 105.3 & 0.09 \\
GX 349+2       & 1.82  & -7.90 & 8.08  & 83216   & 8.08  & -7.90 & 82622  & 64.4  & 0.33 \\
GX 9+1         & 1.37  & 8.36  & -3.77 & 63174   & -3.77 & 8.36  & 64306  & 58.1  & 0.43 \\
GX 17+2        & 1.61  & 14.78 & -7.63 & 72215   & -7.63 & 14.78 & 77127  & 56.3  & 0.26 \\
GX 340+0       & 1.04  & -17.66& 10.90 & 45715   & 10.90 & -17.66& 48782  & 44.3  & 0.35 \\
Cir X-1   & 1.61  & -33.56& 22.10 & 20986   & 22.10 & -33.56& 22005  & 41.2  & 0.60 \\
GRO J1655-40   & 0.82  & -11.63& 10.07 & 32203   & 10.07 & -11.63& 32401  & 36.6  & 0.39 \\
GX 13+1        & 0.80  & 11.65 & -7.05 & 36721   & -7.05 & 11.65 & 36810  & 34.9  & 0.42 \\
4U 1820-303    & 0.74  & -1.77 & -8.21 & 34868   & -8.21 & -1.77 & 33983  & 32.7  & 0.32 \\
GX 3+1         & 0.77  & 2.37  & -0.52 & 35959   & -0.52 & 2.37  & 37159  & 29.5  & 0.11 \\
GRS 1915+105   & 2.12  & 40.75 & -28.04& 7761    & -28.04& 40.75 & 7561   & 24.3  & 0.60 \\
GX 9+9         & 0.68  & 11.95 & 3.39  & 30372   & 3.39  & 11.95 & 31087  & 24.3  & 0.07 \\
4U 1705-440    & 0.43  & -15.53& 6.82  & 18622   & 6.82  & -15.53& 19850  & 19.0  & 0.43 \\
4U 1735-444    & 0.45  & -15.52& 1.23  & 18855   & 1.23  & -15.52& 20718  & 17.3  & 0.69 \\
4U 1636-536    & 0.33  & -25.91& 10.51 & 6652    & 10.51 & -25.91& 8259   & 10.4  & 0.52 \\
GX 354-0       & 0.23  & -4.94 & 2.84  & 11295   & 2.84  & -4.94 & 11326  & 9.9   & 0.27 \\
Ser X-1        & 0.57  & 34.12 & -16.22& 6974    & -16.22& 34.12 & 7842   & 9.7   & 0.37 \\
4U 1630-472    & 0.17  & -19.87& 12.74 & 6550    & 12.74 & -19.87& 7032   & 8.3   & 0.44 \\
GX 339-4       & 0.23  & -20.35& 7.48  & 5068    & 7.48  & -20.35& 5515   & 5.1   & 0.24 \\
\botrule
\end{tabular*}
\footnotetext[1]{Simulated source flux in 2--50~keV [photons $\cdot$ cm$^{-2}$ $\cdot$ s$^{-1}$].}
\footnotetext[2]{Source off-axis angle in fine ($\theta_\mathrm{f}$) and coarse ($\theta_\mathrm{c}$) directions [deg].}
\footnotetext[3]{Camera collected source counts.}
\footnotetext[4]{Combined image source significance.} 
\footnotetext[5]{Combined image source position error [arcmin].} 
\end{table}

\clearpage

\section{Summary and conclusions}\label{sec:conclusion}

In this paper we presented the design and optimization of the coded mask for the \emph{Lunar Electromagnetic Monitor in X-rays} (LEM-X), a proposed all-sky X-ray observatory operating from the Moon's surface which leverages on the heritage of the eXTP/WFM and LOFT/WFM mission concepts. LEM-X is designed to operate in the 2--50~keV energy band, contributing significantly to multi-messenger astrophysics by localizing high-energy transient events and conducting long-term X-ray source monitoring with a large FoV.

The core imaging Unit of LEM-X is a pair of coded-aperture cameras,  based on large-area linear Silicon Drift Detectors (SDDs), each with a sensitive area of 45.5~cm$^2$ and a thickness of 450~$\upmu$m. These detectors, coupled with low-noise front-end electronics, provide excellent spectral and timing performance. The energy resolution is better than 350~eV at 6~keV, and the time resolution reaches 10~$\upmu$s, enabling detailed spectral-timing studies in the 2--50~keV energy band. Spatial resolution is better than 70~$\upmu$m (FWHM) along the anode direction, while the drift direction achieves a resolution of $<$8~mm at 95\% confidence level. These characteristics ensure precise photon localization and spectral-timing measurements, which are critical for coded aperture imaging, transient source detection and characterization.

The optimization of the LEM-X coded mask design has successfully addressed the dual challenge of achieving high scientific performance while guaranteeing mechanical robustness suitable for lunar deployment (Table~\ref{camera_reqs} and Table~\ref{mask_reqs}). The final mask configuration is based on a Modified Uniformly Redundant Array (MURA) code with a cyclic shift applied to enhance mask response symmetry and homogeneity, thus providing a consistent uniform response at the instrument level.
The $260 \times 261~\text{mm}^2$ coded mask has a pitch of 250~$\upmu$m in the fine resolution direction and 15.5~mm in the coarse direction, resulting in a total of 17680 mask elements.
Due to manufacturing and thermo-mechanical constraints, solid opaque ribs were incorporated, leading to  an effective open fraction $f_e = 0.42$.

The imaging performance of the LEM-X camera has been validated through analytical modeling and Monte Carlo simulations. The system point spread function (SPSF) exhibits a single peak with angular dimensions of $4.4'$ in the fine direction and $4.8^\circ$ in the coarse direction, with flat sidelobes indicating low coding noise, a consequence of using a single basic mask pattern to maximize the signal-to-noise ratio and avoid ghost peaks, a crucial design choice given the large FoV and sensitivity requirements.
The Point Source Location Accuracy (PSLA) is better than $0.70'$ in the fine direction and $1^\circ$ in the coarse direction for sources with SNR~$\geq 5$, meeting the mission requirements.

The system performance demonstrated satisfactory margins against the camera-level requirements. 
Sensitivity calculations, benchmarked against the WISEMAN photon-by-photon simulator, show that the instrument achieves a peak sensitivity for an isolated source of $\sim700$~mCrab in 1~s (5$\sigma$), and maintains a sensitivity better than 1000~mCrab across a $\sim45^\circ \times 45^\circ$ wide field-of-view. The effective area exceeds 72~cm$^2$ on-axis, with a field of view at 50\% effective area of approximately $44^\circ \times 44^\circ$, providing a 12--20\% margin over the requirements in Table~\ref{camera_reqs}.

Mechanical analysis confirms that the mask structure, composed of a 150~$\upmu$m thick tungsten sheet under controlled pretension, meets the required stiffness and dimensional stability criteria. A first natural frequency of 120.2~Hz is achieved keeping the in-plane deformation below 10~$\upmu$m in the fine direction and 30~$\upmu$m in the coarse direction. Stress levels reach 412~MPa when the system is subject to random vibrations with GEVS qualification spectrum, still providing a significant safety margin against the tungsten yield strength.

Finally, the imaging capabilities of the full LEM-X Unit (dual-camera configuration) have been demonstrated through simulations of crowded fields such as the Galactic Center. The IROS-based reconstruction pipeline, supported by the \emph{Bloodmoon} and \emph{Darksun} software packages, confirms the instrument’s ability to detect and localize multiple sources with high accuracy and signal-to-noise ratio in a large FoV and validates the overall instrument design described in Section~\ref{sec:instrument_design}.
Considering the effective area and sensitivity maps shown in Figure~\ref{fig:eff_area_sens} and the orientation of the 7 Units composing the instrument, the LEM-X experiment is able to provide a continuous coverage of half of the sky.

The proposed design of the LEM-X camera, based on coded-mask imaging and a SDD-based detection system, offers a robust and high-performance solution for wide-field X-ray imaging from the lunar surface, enabling the mission to contribute significantly to multi-messenger astrophysics through the detection, localization and spectral-timing study of transient high-energy phenomena.

\section*{Acknowledgments}
The authors wish to thank the eXTP/WFM simulation working group for helpful discussions.

The results reported in the article were obtained in the context of the Earth-Moon-Mars (EMM) project, led by INAF in partnership with ASI and CNR. The EMM project is funded under the National Recovery and Resilience Plan (NRRP), Mission 4, Component 2, Investment 3.1: ``Fund for the realisation of an integrated system of research and innovation infrastructures'' - Action 3.1.1 funded by the European Union - NextGenerationEU.

G. D. acknowledges support from INAF RSN-5 mini-grant No. 1.05.24.07.05.
\clearpage

\newpage

\bibliography{bibliography}

@ARTICLE{Abbott2017,
       author = {{Abbott}, B.~P. and {Abbott}, R. and {Abbott}, T.~D. and {Acernese}, F. and {Ackley}, K. and {Adams}, C. and {Adams}, T. and {Addesso}, P. and {Adhikari}, R.~X. and {Adya}, V.~B. and {Affeldt}, C. and {Afrough}, M. and {Agarwal}, B. and {Agathos}, M. and {Agatsuma}, K. and {Aggarwal}, N. and {Aguiar}, O.~D. and {Aiello}, L. and {Ain}, A. and {Ajith}, P. and {Allen}, B. and {Allen}, G. and {Allocca}, A. and {Altin}, P.~A. and {Amato}, A. and {Ananyeva}, A. and {Anderson}, S.~B. and {Anderson}, W.~G. and {Angelova}, S.~V. and {Antier}, S. and {Appert}, S. and {Arai}, K. and {Araya}, M.~C. and {Areeda}, J.~S. and {Arnaud}, N. and {Arun}, K.~G. and {Ascenzi}, S. and {Ashton}, G. and {Ast}, M. and {Aston}, S.~M. and {Astone}, P. and {Atallah}, D.~V. and {Aufmuth}, P. and {Aulbert}, C. and {AultONeal}, K. and {Austin}, C. and {Avila-Alvarez}, A. and {Babak}, S. and {Bacon}, P. and {Bader}, M.~K.~M. and {Bae}, S. and {Bailes}, M. and {Baker}, P.~T. and {Baldaccini}, F. and {Ballardin}, G. and {Ballmer}, S.~W. and {Banagiri}, S. and {Barayoga}, J.~C. and {Barclay}, S.~E. and {Barish}, B.~C. and {Barker}, D. and {Barkett}, K. and {Barone}, F. and {Barr}, B. and {Barsotti}, L. and {Barsuglia}, M. and {Barta}, D. and {Barthelmy}, S.~D. and {Bartlett}, J. and {Bartos}, I. and {Bassiri}, R. and {Basti}, A. and {Batch}, J.~C. and {Bawaj}, M. and {Bayley}, J.~C. and {Bazzan}, M. and {B{\'e}csy}, B. and {Beer}, C. and {Bejger}, M. and {Belahcene}, I. and {Bell}, A.~S. and {Berger}, B.~K. and {Bergmann}, G. and {Bernuzzi}, S. and {Bero}, J.~J. and {Berry}, C.~P.~L. and {Bersanetti}, D. and {Bertolini}, A. and {Betzwieser}, J. and {Bhagwat}, S. and {Bhandare}, R. and {Bilenko}, I.~A. and {Billingsley}, G. and {Billman}, C.~R. and {Birch}, J. and {Birney}, R. and {Birnholtz}, O. and {Biscans}, S. and {Biscoveanu}, S. and {Bisht}, A. and {Bitossi}, M. and {Biwer}, C. and {Bizouard}, M.~A. and {Blackburn}, J.~K. and {Blackman}, J. and {Blair}, C.~D. and {Blair}, D.~G. and {Blair}, R.~M. and {Bloemen}, S. and {Bock}, O. and {Bode}, N. and {Boer}, M. and {Bogaert}, G. and {Bohe}, A. and {Bondu}, F. and {Bonilla}, E. and {Bonnand}, R. and {Boom}, B.~A. and {Bork}, R. and {Boschi}, V. and {Bose}, S. and {Bossie}, K. and {Bouffanais}, Y. and {Bozzi}, A. and {Bradaschia}, C. and {Brady}, P.~R. and {Branchesi}, M. and {Brau}, J.~E. and {Briant}, T. and {Brillet}, A. and {Brinkmann}, M. and {Brisson}, V. and {Brockill}, P. and {Broida}, J.~E. and {Brooks}, A.~F. and {Brown}, D.~A. and {Brown}, D.~D. and {Brunett}, S. and {Buchanan}, C.~C. and {Buikema}, A. and {Bulik}, T. and {Bulten}, H.~J. and {Buonanno}, A. and {Buskulic}, D. and {Buy}, C. and {Byer}, R.~L. and {Cabero}, M. and {Cadonati}, L. and {Cagnoli}, G. and {Cahillane}, C. and {Calder{\'o}n Bustillo}, J. and {Callister}, T.~A. and {Calloni}, E. and {Camp}, J.~B. and {Canepa}, M. and {Canizares}, P. and {Cannon}, K.~C. and {Cao}, H. and {Cao}, J. and {Capano}, C.~D. and {Capocasa}, E. and {Carbognani}, F. and {Caride}, S. and {Carney}, M.~F. and {Carullo}, G. and {Casanueva Diaz}, J. and {Casentini}, C. and {Caudill}, S. and {Cavagli{\`a}}, M. and {Cavalier}, F. and {Cavalieri}, R. and {Cella}, G. and {Cepeda}, C.~B. and {Cerd{\'a}-Dur{\'a}n}, P. and {Cerretani}, G. and {Cesarini}, E. and {Chamberlin}, S.~J. and {Chan}, M. and {Chao}, S. and {Charlton}, P. and {Chase}, E. and {Chassande-Mottin}, E. and {Chatterjee}, D. and {Chatziioannou}, K. and {Cheeseboro}, B.~D. and {Chen}, H.~Y. and {Chen}, X. and {Chen}, Y. and {Cheng}, H. -P. and {Chia}, H. and {Chincarini}, A. and {Chiummo}, A. and {Chmiel}, T. and {Cho}, H.~S. and {Cho}, M. and {Chow}, J.~H. and {Christensen}, N. and {Chu}, Q. and {Chua}, A.~J.~K. and {Chua}, S.},
        title = "{GW170817: Observation of Gravitational Waves from a Binary Neutron Star Inspiral}",
      journal = {\prl},
         year = 2017,
        month = oct,
       volume = {119},
       number = {16},
          eid = {161101},
        pages = {161101},
          doi = {10.1103/PhysRevLett.119.161101},
       adsurl = {https://ui.adsabs.harvard.edu/abs/2017PhRvL.119p1101A}
}

@ARTICLE{Troja2017,
       author = {{Troja}, E. and {Piro}, L. and {van Eerten}, H. and {Wollaeger}, R.~T. and {Im}, M. and {Fox}, O.~D. and {Butler}, N.~R. and {Cenko}, S.~B. and {Sakamoto}, T. and {Fryer}, C.~L. and {Ricci}, R. and {Lien}, A. and {Ryan}, R.~E. and {Korobkin}, O. and {Lee}, S. -K. and {Burgess}, J.~M. and {Lee}, W.~H. and {Watson}, A.~M. and {Choi}, C. and {Covino}, S. and {D'Avanzo}, P. and {Fontes}, C.~J. and {Gonz{\'a}lez}, J. Becerra and {Khandrika}, H.~G. and {Kim}, J. and {Kim}, S. -L. and {Lee}, C. -U. and {Lee}, H.~M. and {Kutyrev}, A. and {Lim}, G. and {S{\'a}nchez-Ram{\'\i}rez}, R. and {Veilleux}, S. and {Wieringa}, M.~H. and {Yoon}, Y.},
        title = "{The X-ray counterpart to the gravitational-wave event GW170817}",
      journal = {\nat},
         year = 2017,
        month = nov,
       volume = {551},
       number = {7678},
        pages = {71-74},
          doi = {10.1038/nature24290},
       adsurl = {https://ui.adsabs.harvard.edu/abs/2017Natur.551...71T}
}

@ARTICLE{Goldstein2017,
       author = {{Goldstein}, A. and {Veres}, P. and {Burns}, E. and {Briggs}, M.~S. and {Hamburg}, R. and {Kocevski}, D. and {Wilson-Hodge}, C.~A. and {Preece}, R.~D. and {Poolakkil}, S. and {Roberts}, O.~J. and {Hui}, C.~M. and {Connaughton}, V. and {Racusin}, J. and {von Kienlin}, A. and {Dal Canton}, T. and {Christensen}, N. and {Littenberg}, T. and {Siellez}, K. and {Blackburn}, L. and {Broida}, J. and {Bissaldi}, E. and {Cleveland}, W.~H. and {Gibby}, M.~H. and {Giles}, M.~M. and {Kippen}, R.~M. and {McBreen}, S. and {McEnery}, J. and {Meegan}, C.~A. and {Paciesas}, W.~S. and {Stanbro}, M.},
        title = "{An Ordinary Short Gamma-Ray Burst with Extraordinary Implications: Fermi-GBM Detection of GRB 170817A}",
      journal = {\apjl},
         year = 2017,
        month = oct,
       volume = {848},
       number = {2},
          eid = {L14},
        pages = {L14},
          doi = {10.3847/2041-8213/aa8f41},
       adsurl = {https://ui.adsabs.harvard.edu/abs/2017ApJ...848L..14G}
}

@ARTICLE{IceCube2013,
       author = {{IceCube Collaboration}},
        title = "{Evidence for High-Energy Extraterrestrial Neutrinos at the IceCube Detector}",
      journal = {Science},
         year = 2013,
        month = nov,
       volume = {342},
       number = {6161},
          eid = {1242856},
        pages = {1242856},
          doi = {10.1126/science.1242856},
       adsurl = {https://ui.adsabs.harvard.edu/abs/2013Sci...342E...1I}
}

@INPROCEEDINGS{Nuti2024,
       author = {{Nuti}, Alessio and {Dilillo}, Giuseppe and {Ceraudo}, Francesco and {Della Casa}, Giovanni and {Del Monte}, Ettore and {Evangelista}, Yuri and {Feroci}, Marco and {Lombardi}, Giovanni and {Rapisarda}, Massimo and {Esposito}, Francesca and {Donnarumma}, Immacolata and {Cortesi}, Ugo and {D'Amico}, Fabio and {Turchi}, Alessandro and {Gai}, Marco and {Argan}, Andrea},
        title = "{The Lunar Electromagnetic Monitor in X-rays (LEM-X): optimization of the instrument layout and trade-off study for the observatory location on the Moon surface}",
    booktitle = {Space Telescopes and Instrumentation 2024: Ultraviolet to Gamma Ray},
         year = 2024,
       editor = {{den Herder}, Jan-Willem A. and {Nikzad}, Shouleh and {Nakazawa}, Kazuhiro},
       series = {Society of Photo-Optical Instrumentation Engineers (SPIE) Conference Series},
       volume = {13093},
        month = aug,
          eid = {130937O},
        pages = {130937O},
          doi = {10.1117/12.3030692},
       adsurl = {https://ui.adsabs.harvard.edu/abs/2024SPIE13093E..7ON},
}

@INPROCEEDINGS{Delmonte2024,
       author = {{Del Monte}, Ettore and {Ceraudo}, Francesco and {Della Casa}, Giovanni and {Dilillo}, Giuseppe and {Evangelista}, Yuri and {Feroci}, Marco and {Nuti}, Alessio and {Rapisarda}, Massimo and {Bertuccio}, Giuseppe and {Campana}, Riccardo and {Demenev}, Evgeny and {Ficorella}, Francesco and {Fiorini}, Mauro and {Grassi}, Marco and {Malcovati}, Piero and {Mele}, Filippo and {Pepponi}, Giancarlo and {Zampa}, Gianluigi and {Esposito}, Francesca and {Donnarumma}, Immacolata and {Cortesi}, Ugo and {D'Amico}, Fabio and {Turchi}, Alessandro and {Gai}, Marco and {Argan}, Andrea},
        title = "{Status of the Lunar Electromagnetic Monitor in X-rays (LEM-X)}",
    booktitle = {Space Telescopes and Instrumentation 2024: Ultraviolet to Gamma Ray},
         year = 2024,
       editor = {{den Herder}, Jan-Willem A. and {Nikzad}, Shouleh and {Nakazawa}, Kazuhiro},
       series = {Society of Photo-Optical Instrumentation Engineers (SPIE) Conference Series},
       volume = {13093},
        month = aug,
          eid = {130931U},
        pages = {130931U},
          doi = {10.1117/12.3018838},
       adsurl = {https://ui.adsabs.harvard.edu/abs/2024SPIE13093E..1UD}
}

@INPROCEEDINGS{Hernanz2024,
       author = {{Hernanz}, Margarita and {Feroci}, Marco and {Evangelista}, Yuri and {Meuris}, Aline and {Schanne}, St{\'e}phane and {Zampa}, Gianluigi and {Tenzer}, Chris and {Bayer}, J{\"o}rg and {Nowosielski}, Witold and {Michalska}, Malgorzata and {Kalemci}, Emrah and {Sungur}, M{\"u}berra and {Brandt}, S{\o}ren and {Kuvvetli}, Irfan and {Alvarez Franco}, Daniel and {Carmona}, Alex and {G{\'a}lvez}, Jos{\'e}-Luis and {Patruno}, Alessandro and {In't Zand}, Jean and {Zwart}, Frans and {Santangelo}, Andrea and {Bozzo}, Enrico and {Zhang}, Shuang-Nan and {Lu}, Fangjun and {Xu}, Yupeng and {Campana}, Riccardo and {Del Monte}, Ettore and {Ceraudo}, Francesco and {Nuti}, Alessio and {Della Casa}, Giovanni and {Argan}, Andrea and {Minervini}, Gabriele and {Antonelli}, Matias and {Bonvicini}, Valter and {Boezio}, Mirko and {Cirrincione}, Daniela and {Munini}, Riccardo and {Rachevski}, Alexandre and {Vacchi}, Andrea and {Zampa}, Nicola and {Rashevskaya}, Irina and {Ficorella}, Francesco and {Picciotto}, Antonino and {Zorzi}, Nicola and {Baudin}, David and {Bouyjou}, Florent and {Gevin}, Olivier and {Limousin}, Olivier and {Hedderman}, Paul and {Pliego}, Samuel and {Xiong}, Hao and {de la Rie}, Rob and {Laubert}, Phillip and {Aitink-Kroes}, Gabby and {Kuiper}, Lucien and {Orleanski}, Piotr and {Skup}, Konrad and {Tcherniak}, Denis and {Turhan}, Onur and {Bozkurt}, Ayhan and {Onat}, Ahmet},
        title = "{The Wide Field Monitor (WFM) of the China-Europe eXTP (enhanced X-ray Timing and Polarimetry) mission}",
    booktitle = {Space Telescopes and Instrumentation 2024: Ultraviolet to Gamma Ray},
         year = 2024,
       editor = {{den Herder}, Jan-Willem A. and {Nikzad}, Shouleh and {Nakazawa}, Kazuhiro},
       series = {Society of Photo-Optical Instrumentation Engineers (SPIE) Conference Series},
       volume = {13093},
        month = aug,
          eid = {130931Y},
        pages = {130931Y},
          doi = {10.1117/12.3020020},
archivePrefix = {arXiv},
       eprint = {2411.03050},
 primaryClass = {astro-ph.IM},
       adsurl = {https://ui.adsabs.harvard.edu/abs/2024SPIE13093E..1YH}
}

@ARTICLE{Rachevski2014J,
       author = {{Rachevski}, A. and {Zampa}, G. and {Zampa}, N. and {Campana}, R. and {Evangelista}, Y. and {Giacomini}, G. and {Picciotto}, A. and {Bellutti}, P. and {Feroci}, M. and {Labanti}, C. and {Piemonte}, C. and {Vacchi}, A.},
        title = "{Large-area linear Silicon Drift Detector design for X-ray experiments}",
      journal = {Journal of Instrumentation},
         year = 2014,
        month = jul,
       volume = {9},
       number = {7},
          eid = {P07014},
        pages = {P07014},
          doi = {10.1088/1748-0221/9/07/P07014},
       adsurl = {https://ui.adsabs.harvard.edu/abs/2014JInst...9P7014R}
}

@INPROCEEDINGS{Mele2025a,
  author = {Mele, Filippo and Grassi, Marco and Dedolli, Irisa and Malcovati, Piero and Campana, Riccardo and Del Monte, Ettore and Evangelista, Yuri and Feroci, Marco and Bertuccio, Giuseppe},
  title = {{The Front-End Charge Readout IC for the LEM-X Mission Concept}},
  booktitle = "Proceedings of SIE 2024",
editor="Valle, Maurizio and Gastaldo, Paolo and Limiti, Ernesto",
  year = "2025",
  publisher = "Springer Nature Switzerland",
  pages = "170--176",
  isbn = "978-3-031-71518-1"
}

@INPROCEEDINGS{Ceraudo2024SDD,
       author = {{Ceraudo}, Francesco and {Della Casa}, Giovanni and {Bertuccio}, Giuseppe and {Bonvicini}, Walter and {Campana}, Riccardo and {Cirrincione}, Daniela and {Del Monte}, Ettore and {Evangelista}, Yuri and {Feroci}, Marco and {Ficorella}, Francesco and {Fiorini}, Mauro and {Grassi}, Marco and {Malcovati}, Piero and {Mele}, Filippo and {Pepponi}, Giancarlo and {Rachevski}, Alexandre and {Rashevskaya}, Irina and {Zampa}, Gianluigi and {Zampa}, Nicola and {Zorzi}, Nicola},
        title = "{Imaging and spectroscopic performances of the silicon drift detector of the wide field monitor}",
    booktitle = {Space Telescopes and Instrumentation 2024: Ultraviolet to Gamma Ray},
       editor = {{den Herder}, Jan-Willem A. and {Nikzad}, Shouleh and {Nakazawa}, Kazuhiro},
         year = 2024,
       series = {Society of Photo-Optical Instrumentation Engineers (SPIE) Conference Series},
       volume = {13093},
        month = aug,
          eid = {130936U},
        pages = {130936U},
          doi = {10.1117/12.3020204},
       adsurl = {https://ui.adsabs.harvard.edu/abs/2024SPIE13093E..6UC}
}

@ARTICLE{Campana2011,
       author = {{Campana}, R. and {Zampa}, G. and {Feroci}, M. and {Vacchi}, A. and {Bonvicini}, V. and {Del Monte}, E. and {Evangelista}, Y. and {Fuschino}, F. and {Labanti}, C. and {Marisaldi}, M. and {Muleri}, F. and {Pacciani}, L. and {Rapisarda}, M. and {Rashevsky}, A. and {Rubini}, A. and {Soffitta}, P. and {Zampa}, N. and {Baldazzi}, G. and {Costa}, E. and {Donnarumma}, I. and {Grassi}, M. and {Lazzarotto}, F. and {Malcovati}, P. and {Mastropietro}, M. and {Morelli}, E. and {Picolli}, L.},
        title = "{Imaging performance of a large-area Silicon Drift Detector for X-ray astronomy}",
      journal = {Nuclear Instruments and Methods in Physics Research A},
         year = 2011,
        month = mar,
       volume = {633},
       number = {1},
        pages = {22-30},
          doi = {10.1016/j.nima.2010.12.237},
       adsurl = {https://ui.adsabs.harvard.edu/abs/2011NIMPA.633...22C}
}

@INPROCEEDINGS{Zwart2022,
       author = {{Zwart}, F. and {Tacken}, R. and {in't Zand}, J.~J.~M. and {de la Rie}, R. and {Limpens}, M. and {Kochanowski}, C. and {Aitink-Kroes}, G. and {van Baren}, C. and {Bayer}, J. and {Baudin}, D. and {Ceraudo}, F. and {Evangelista}, Y. and {Feroci}, M. and {Frericks}, M. and {G{\'a}lvez}, J. -L. and {Gevin}, O. and {Hernanz}, M. and {Hormaetxe}, A. and {Laubert}, P. and {Meuris}, A. and {Nab}, J. and {Neelis}, J. and {Tenzer}, C. and {Vogel}, C. and {Zampa}, G.},
        title = "{The detector/readout-electronics assembly of the eXTP wide field monitor}",
    booktitle = {Space Telescopes and Instrumentation 2022: Ultraviolet to Gamma Ray},
       editor = {{den Herder}, Jan-Willem A. and {Nikzad}, Shouleh and {Nakazawa}, Kazuhiro},
         year = 2022,
       series = {Society of Photo-Optical Instrumentation Engineers (SPIE) Conference Series},
       volume = {12181},
        month = aug,
          eid = {1218167},
        pages = {1218167},
          doi = {10.1117/12.2629406},
       adsurl = {https://ui.adsabs.harvard.edu/abs/2022SPIE12181E..67Z}
}

@INPROCEEDINGS{Evangelista2014,
       author = {{Evangelista}, Y. and {Donnarumma}, I. and {Campana}, R. and {Schmid}, C. and {Feroci}, M.},
        title = "{Instrumental and scientific simulations of the LOFT wide field monitor}",
    booktitle = {Space Telescopes and Instrumentation 2014: Ultraviolet to Gamma Ray},
       editor = {{Takahashi}, Tadayuki and {den Herder}, Jan-Willem A. and {Bautz}, Mark},
         year = 2014,
       series = {Society of Photo-Optical Instrumentation Engineers (SPIE) Conference Series},
       volume = {9144},
        month = jul,
          eid = {914468},
        pages = {914468},
          doi = {10.1117/12.2055772},
       adsurl = {https://ui.adsabs.harvard.edu/abs/2014SPIE.9144E..68E}
}

@INPROCEEDINGS{Evangelista2012,
       author = {{Evangelista}, Y. and {Campana}, R. and {Del Monte}, E. and {Donnarumma}, I. and {Feroci}, M. and {Muleri}, F. and {Pacciani}, L. and {Soffitta}, P. and {Rachevski}, A. and {Vacchi}, A. and {Zampa}, G. and {Zampa}, N. and {Suchy}, S. and {Brandt}, S. and {Budtz-J{\o}rgensen}, C. and {Hernanz}, M.},
        title = "{Simulations of the x-ray imaging capabilities of the silicon drift detectors (SDD) for the LOFT wide-field monitor}",
    booktitle = {Space Telescopes and Instrumentation 2012: Ultraviolet to Gamma Ray},
       editor = {{Takahashi}, Tadayuki and {Murray}, Stephen S. and {den Herder}, Jan-Willem A.},
         year = 2012,
       series = {Society of Photo-Optical Instrumentation Engineers (SPIE) Conference Series},
       volume = {8443},
        month = sep,
          eid = {84435P},
        pages = {84435P},
          doi = {10.1117/12.926000},
       adsurl = {https://ui.adsabs.harvard.edu/abs/2012SPIE.8443E..5PE}
}

@ARTICLE{GottesmanFenimore1989,
       author = {{Gottesman}, Stephen R. and {Fenimore}, E.~E.},
        title = "{New family of binary arrays for coded aperture imaging}",
      journal = {\ao},
         year = 1989,
        month = oct,
       volume = {28},
       number = {20},
        pages = {4344-4352},
          doi = {10.1364/AO.28.004344},
       adsurl = {https://ui.adsabs.harvard.edu/abs/1989ApOpt..28.4344G}
}

@ARTICLE{Caroli1987,
       author = {{Caroli}, E. and {Stephen}, J.~B. and {Di Cocco}, G. and {Natalucci}, L. and {Spizzichino}, A.},
        title = "{Coded Aperture Imaging in X-Ray and Gamma-Ray Astronomy}",
      journal = {\ssr},
         year = 1987,
        month = sep,
       volume = {45},
       number = {3-4},
        pages = {349-403},
          doi = {10.1007/BF00171998},
       adsurl = {https://ui.adsabs.harvard.edu/abs/1987SSRv...45..349C}
}

@PHDTHESIS{intZand1992,
       author = {{in 't Zand}, Johannes Joseph Marie},
        title = "{A coded-mask imager as monitor of Galactic X-ray sources}",
       school = {Netherlands Institute for Space Research},
         year = 1992,
        month = oct,
       adsurl = {https://ui.adsabs.harvard.edu/abs/1992PhDT........16I}
}

@PHDTHESIS{Accorsi2001,
       author = {{Accorsi}, Roberto},
        title = "{Design of a near-field coded aperture cameras for high-resolution medical and industrial gamma-ray imaging}",
       school = {Massachusetts Institute of Technology. Dept. of Nuclear Engineering},
         year = 2001,
        month = jul,
       url = {http://hdl.handle.net/1721.1/8684}
}

@ARTICLE{Skinner2008,
       author = {{Skinner}, Gerald K.},
        title = "{Sensitivity of coded mask telescopes}",
      journal = {\ao},
         year = 2008,
        month = may,
       volume = {47},
       number = {15},
        pages = {2739-2749},
          doi = {10.1364/AO.47.002739},
       adsurl = {https://ui.adsabs.harvard.edu/abs/2008ApOpt..47.2739S}
}

@ARTICLE{Braga2020,
       author = {{Braga}, Jo{\~a}o},
        title = "{Coded Aperture Imaging in High-energy Astrophysics}",
      journal = {\pasp},
         year = 2020,
        month = jan,
       volume = {132},
       number = {1007},
          eid = {012001},
        pages = {012001},
          doi = {10.1088/1538-3873/ab450a},
       adsurl = {https://ui.adsabs.harvard.edu/abs/2020PASP..132a2001B}
}

@INCOLLECTION{Goldwurm2022,
       author = {{Goldwurm}, Andrea and {Gros}, Aleksandra},
        title = "{Coded Mask Instruments for Gamma-Ray Astronomy}",
    booktitle = {Handbook of X-ray and Gamma-ray Astrophysics},
       editor = {{Bambi}, Cosimo and {Sangangelo}, Andrea},
    publisher ="Springer Nature Singapore",
         year = 2022,
          eid = {15},
        pages = {15},
          doi = {10.1007/978-981-16-4544-0_44-1},
       adsurl = {https://ui.adsabs.harvard.edu/abs/2022hxga.book...15G}
}

@ARTICLE{FenimoreCannon1978,
       author = {{Fenimore}, E.~E. and {Cannon}, T.~M.},
        title = "{Coded aperture imaging with uniformly redundant arrays}",
      journal = {\ao},
         year = 1978,
        month = feb,
       volume = {17},
        pages = {337-347},
          doi = {10.1364/AO.17.000337},
       adsurl = {https://ui.adsabs.harvard.edu/abs/1978ApOpt..17..337F}
}

@ARTICLE{Goldwurm1995,
       author = {{Goldwurm}, A.},
        title = "{Imaging Techniques Applied to the Coded Mask SIGMA Telescope}",
      journal = {Experimental Astronomy},
         year = 1995,
        month = dec,
       volume = {6},
       number = {4},
        pages = {9-18},
          doi = {10.1007/BF00419253},
       adsurl = {https://ui.adsabs.harvard.edu/abs/1995ExA.....6....9G}
}

@ARTICLE{SkinnerPonman1994,
       author = {{Skinner}, G.~K. and {Ponman}, T.~J.},
        title = "{On the Properties of Images from Coded Mask Telescopes}",
      journal = {\mnras},
         year = 1994,
        month = apr,
       volume = {267},
        pages = {518},
          doi = {10.1093/mnras/267.3.518},
       adsurl = {https://ui.adsabs.harvard.edu/abs/1994MNRAS.267..518S}
}

@BOOK{Baumert1971,
    author = {Baumert, Leonard D.},
    isbn = {9783540364535},
    language = {eng},
    publisher = {Springer-Verlag},
    series = {Lecture notes in mathematics, 182},
    title = {Cyclic difference sets / Leonard D. Baumert.},
    url = {https://link.springer.com/10.1007/BFb0061260},
    year = {1971},
    address = {Berlin ;},
    booktitle = {Cyclic difference sets},
}

@ARTICLE{Feroci2007,
       author = {{Feroci}, M. and {Costa}, E. and {Soffitta}, P. and {Del Monte}, E. and {di Persio}, G. and {Donnarumma}, I. and {Evangelista}, Y. and {Frutti}, M. and {Lapshov}, I. and {Lazzarotto}, F. and {Mastropietro}, M. and {Morelli}, E. and {Pacciani}, L. and {Porrovecchio}, G. and {Rapisarda}, M. and {Rubini}, A. and {Tavani}, M. and {Argan}, A.},
        title = "{SuperAGILE: The hard X-ray imager for the AGILE space mission}",
      journal = {Nuclear Instruments and Methods in Physics Research A},
         year = 2007,
        month = nov,
       volume = {581},
       number = {3},
        pages = {728-754},
          doi = {10.1016/j.nima.2007.07.147},
       adsurl = {https://ui.adsabs.harvard.edu/abs/2007NIMPA.581..728F}
}

@ARTICLE{Feroci2010,
       author = {{Feroci}, M. and {Costa}, E. and {Del Monte}, E. and {Donnarumma}, I. and {Evangelista}, Y. and {Lapshov}, I. and {Lazzarotto}, F. and {Pacciani}, L. and {Rapisarda}, M. and {Soffitta}, P. and {di Persio}, G. and {Frutti}, M. and {Mastropietro}, M. and {Morelli}, E. and {Porrovecchio}, G. and {Rubini}, A. and {Antonelli}, A. and {Argan}, A. and {Barbiellini}, G. and {Boffelli}, F. and {Bulgarelli}, A. and {Caraveo}, P. and {Cattaneo}, P.~W. and {Chen}, A.~W. and {Cocco}, V. and {Colafrancesco}, S. and {Cutini}, S. and {D'Ammando}, F. and {de Paris}, G. and {Di Cocco}, G. and {Fanari}, G. and {Ferrari}, A. and {Fiorini}, M. and {Fornari}, F. and {Fuschino}, F. and {Froysland}, T. and {Galli}, M. and {Gasparrini}, D. and {Gianotti}, F. and {Giommi}, P. and {Giuliani}, A. and {Labanti}, C. and {Liello}, F. and {Lipari}, P. and {Longo}, F. and {Mattaini}, E. and {Marisaldi}, M. and {Mauri}, A. and {Mauri}, F. and {Mereghetti}, S. and {Moretti}, E. and {Morselli}, A. and {Pellizzoni}, A. and {Perotti}, F. and {Piano}, G. and {Picozza}, P. and {Pilia}, M. and {Pittori}, C. and {Pontoni}, C. and {Preger}, B. and {Prest}, M. and {Primavera}, R. and {Pucella}, G. and {Rappoldi}, A. and {Rossi}, E. and {Sabatini}, S. and {Santolamazza}, P. and {Tavani}, M. and {Stellato}, S. and {Tamburelli}, F. and {Traci}, A. and {Trifoglio}, M. and {Trois}, A. and {Vallazza}, E. and {Vercellone}, S. and {Verrecchia}, F. and {Vittorini}, V. and {Zambra}, A. and {Zanello}, D. and {Salotti}, L.},
        title = "{Monitoring the hard X-ray sky with SuperAGILE}",
      journal = {\aap},
         year = 2010,
        month = feb,
       volume = {510},
          eid = {A9},
        pages = {A9},
          doi = {10.1051/0004-6361/200912972},
       adsurl = {https://ui.adsabs.harvard.edu/abs/2010A&A...510A...9F}
}

@ARTICLE{Busboom1998,
       author = {{Busboom}, A. and {Elders-Boll}, H. and {Schotten}, H.~D.},
        title = "{Uniformly Redundant Arrays}",
      journal = {Experimental Astronomy},
         year = 1998,
        month = jan,
       volume = {8},
       number = {2},
        pages = {97-123},
          doi = {10.1023/A:1007966830741},
       adsurl = {https://ui.adsabs.harvard.edu/abs/1998ExA.....8...97B}
}

@ARTICLE{intZand1994,
       author = {{in 't Zand}, J.~J.~M. and {Heise}, J. and {Jager}, R.},
        title = "{The optimum open fraction of coded apertures. With an application to the wide field X-ray cameras of SAX}",
      journal = {\aap},
         year = 1994,
        month = aug,
       volume = {288},
        pages = {665-674},
       adsurl = {https://ui.adsabs.harvard.edu/abs/1994A&A...288..665I}
}

@ARTICLE{Shutler2013,
       author = {{Shutler}, Paul M.~E. and {Springham}, Stuart V. and {Talebitaher}, Alireza},
        title = "{Mask design and fabrication in coded aperture imaging}",
      journal = {Nuclear Instruments and Methods in Physics Research A},
         year = 2013,
        month = may,
       volume = {709},
        pages = {129-142},
          doi = {10.1016/j.nima.2013.01.032},
       adsurl = {https://ui.adsabs.harvard.edu/abs/2013NIMPA.709..129S}
}

@INPROCEEDINGS{Ceraudo2024wiseman,
       author = {{Ceraudo}, Francesco and {Evangelista}, Yuri and {Hernanz}, Margarita and {in't Zand}, Jean and {Kuiper}, Lucien and {Patruno}, Alessandro},
        title = "{Development of the end-to-end simulator of the WFM camera}",
    booktitle = {Space Telescopes and Instrumentation 2024: Ultraviolet to Gamma Ray},
       editor = {{den Herder}, Jan-Willem A. and {Nikzad}, Shouleh and {Nakazawa}, Kazuhiro},
         year = 2024,
       series = {Society of Photo-Optical Instrumentation Engineers (SPIE) Conference Series},
       volume = {13093},
        month = aug,
          eid = {130936T},
        pages = {130936T},
          doi = {10.1117/12.3020083},
       adsurl = {https://ui.adsabs.harvard.edu/abs/2024SPIE13093E..6TC}
}

@ARTICLE{Hammersley1992,
       author = {{Hammersley}, Andrew and {Ponman}, Trevor and {Skinner}, Gerry},
        title = "{Reconstruction of images from a coded-aperture box camera}",
      journal = {Nuclear Instruments and Methods in Physics Research A},
         year = 1992,
        month = jan,
       volume = {311},
       number = {3},
        pages = {585-594},
          doi = {10.1016/0168-9002(92)90659-R},
       adsurl = {https://ui.adsabs.harvard.edu/abs/1992NIMPA.311..585H}
}

@ARTICLE{Giancarli2025,
       author = {{Giancarli}, Edoardo and {Dilillo}, Giuseppe},
        title = "{Development of the Iterative Removal of Sources (IROS) pipeline for the LEM-X experiment}",
      journal = {in preparation},
         year = 2025
}

@ARTICLE{DelMonte2025,
  author = {{Del Monte}, Ettore and {Evangelista}, Yuri and {Feroci}, Marco and {Ceraudo}, Francesco and {Della Casa}, Giovanni and {Nuti}, Alessio},
  title = "{The Lunar Electromagnetic Monitor in X-rays (LEM-X)}",
  journal = {in preparation},
  year = 2025
}

@ARTICLE{Proctor1979,
       author = {{Proctor}, R.~J. and {Skinner}, G.~K. and {Willmore}, A.~P.},
        title = "{The design of optimum coded mask X-ray telescopes.}",
      journal = {\mnras},
         year = 1979,
        month = jun,
       volume = {187},
        pages = {633-643},
          doi = {10.1093/mnras/187.3.633},
       adsurl = {https://ui.adsabs.harvard.edu/abs/1979MNRAS.187..633P}
}

@ARTICLE{Levine1996,
       author = {{Levine}, Alan M. and {Bradt}, Hale and {Cui}, Wei and {Jernigan}, J.~G. and {Morgan}, Edward H. and {Remillard}, Ronald and {Shirey}, Robert E. and {Smith}, Donald A.},
        title = "{First Results from the All-Sky Monitor on the Rossi X-Ray Timing Explorer}",
      journal = {\apjl},
         year = 1996,
        month = sep,
       volume = {469},
        pages = {L33},
          doi = {10.1086/310260},
archivePrefix = {arXiv},
       eprint = {astro-ph/9608109},
 primaryClass = {astro-ph},
       adsurl = {https://ui.adsabs.harvard.edu/abs/1996ApJ...469L..33L}
}

@INPROCEEDINGS{Hernanz2018,
       author = {{Hernanz}, M. and {Brandt}, S. and {Feroci}, M. and {Orleansky}, P. and {Santangelo}, A. and {Schanne}, S. and {Wu}, Xin and {in't Zand}, J. and {Zhang}, S.~N. and {Xu}, Y.~P. and {Bozzo}, E. and {Evangelista}, Y. and {G{\'a}lvez}, J.~L. and {Tenzer}, C. and {Zwart}, F. and {Lu}, F.~J. and {Zhang}, S. and {Cheng}, T.~X. and {Ambrosino}, F. and {Argan}, A. and {Del Monte}, E. and {Budtz-Jorgensen}, C. and {Lund}, N. and {Olsen}, P. and {Mansanet}, C. and {Campana}, R. and {Fuschino}, F. and {Labanti}, C. and {Rachevski}, A. and {Vacchi}, A. and {Zampa}, G. and {Zampa}, N. and {Rashevskaya}, I. and {Bellutti}, P. and {Borghi}, G. and {Ficorella}, F. and {Picciotto}, A. and {Zorzi}, N. and {Limousin}, O. and {Meris}, A.},
        title = "{The wide field monitor onboard the eXTP mission}",
    booktitle = {Space Telescopes and Instrumentation 2018: Ultraviolet to Gamma Ray},
         year = 2018,
       editor = {{den Herder}, Jan-Willem A. and {Nikzad}, Shouleh and {Nakazawa}, Kazuhiro},
       series = {Society of Photo-Optical Instrumentation Engineers (SPIE) Conference Series},
       volume = {10699},
        month = jul,
          eid = {1069948},
        pages = {1069948},
          doi = {10.1117/12.2313214},
archivePrefix = {arXiv},
       eprint = {1807.09330},
 primaryClass = {astro-ph.IM},
       adsurl = {https://ui.adsabs.harvard.edu/abs/2018SPIE10699E..48H}
}

@INPROCEEDINGS{Brandt2014,
       author = {{Brandt}, S. and {Hernanz}, M. and {Alvarez}, L. and {Argan}, A. and {Artigues}, B. and {Azzarello}, P. and {Barret}, D. and {Bozzo}, E. and {Budtz-J{\o}rgensen}, C. and {Campana}, R. and {Cros}, A. and {del Monte}, E. and {Donnarumma}, I. and {Evangelista}, Y. and {Feroci}, M. and {Galvez Sanchez}, J.~L. and {G{\"o}tz}, D. and {Hansen}, F. and {den Herder}, J.~W. and {Hudec}, R. and {Huovelin}, J. and {Karelin}, D. and {Korpela}, S. and {Lund}, N. and {Michalska}, M. and {Olsen}, P. and {Orleanski}, P. and {Pedersen}, S. and {Pohl}, M. and {Rachevski}, A. and {Santangelo}, A. and {Schanne}, S. and {Schmid}, C. and {Suchy}, S. and {Tenzer}, C. and {Vacchi}, A. and {Walton}, D. and {Wilms}, J. and {Zampa}, G. and {Zampa}, N. and {in't Zand}, J. and {Zane}, S. and {Zdziarski}, A. and {Zwart}, F.},
        title = "{The design of the wide field monitor for the LOFT mission}",
    booktitle = {Space Telescopes and Instrumentation 2014: Ultraviolet to Gamma Ray},
         year = 2014,
       editor = {{Takahashi}, Tadayuki and {den Herder}, Jan-Willem A. and {Bautz}, Mark},
       series = {Society of Photo-Optical Instrumentation Engineers (SPIE) Conference Series},
       volume = {9144},
        month = jul,
          eid = {91442V},
        pages = {91442V},
          doi = {10.1117/12.2055885},
archivePrefix = {arXiv},
       eprint = {1408.6540},
 primaryClass = {astro-ph.IM},
       adsurl = {https://ui.adsabs.harvard.edu/abs/2014SPIE.9144E..2VB}
}

@ARTICLE{Ajello2008,
       author = {{Ajello}, M. and {Greiner}, J. and {Sato}, G. and {Willis}, D.~R. and {Kanbach}, G. and {Strong}, A.~W. and {Diehl}, R. and {Hasinger}, G. and {Gehrels}, N. and {Markwardt}, C.~B. and {Tueller}, J.},
        title = "{Cosmic X-Ray Background and Earth Albedo Spectra with Swift BAT}",
      journal = {\apj},
         year = 2008,
        month = dec,
       volume = {689},
       number = {2},
        pages = {666-677},
          doi = {10.1086/592595},
archivePrefix = {arXiv},
       eprint = {0808.3377},
 primaryClass = {astro-ph},
       adsurl = {https://ui.adsabs.harvard.edu/abs/2008ApJ...689..666A}
}

\end{document}